\documentclass[12pt,a4paper]{article} 
\usepackage{graphicx}
\usepackage[T1]{fontenc}
\usepackage[utf8]{inputenc}
\usepackage{amsmath}
\usepackage{amssymb}
\usepackage{mathtools}
\usepackage{dsfont}
\usepackage{slashed}
\usepackage{siunitx}
\usepackage{bbold}
\usepackage{color}
\usepackage[dvipsnames]{xcolor}
\usepackage[backend=bibtex,style=phys,sorting=none,citestyle=numeric-comp,eprint=true]{biblatex}
\usepackage[
colorlinks=true,
allcolors=blue
]{hyperref}
\usepackage{cleveref}
\usepackage{subfigure}
\usepackage{placeins}
\newcommand{\beq}{\begin{equation}}
\newcommand{\eeq}{\end{equation}}
\newcommand{\bea}{\begin{eqnarray}}
\newcommand{\eea}{\end{eqnarray}}
\newcommand{\x}{{\bf x}}

\newcommand{\One}{1\kern-4.5pt1}

\graphicspath{{figures/},{inkscape/},{data/}}
\newcommand{\im}{i}

\DeclareMathOperator{\atan}{atan2}

\newcommand{\matr}[2]{\left(\begin{array}{#1}#2\end{array}\right)}
\newcommand{\pdagger}{{\phantom{\dagger}}}
\newcommand{\trans}{\ensuremath{\mathsf{T}}}

\newcommand{\erwartung}[1]{\ensuremath{\left\langle#1\right\rangle}}
\newcommand{\br}[1]{\ensuremath{\left(#1\right)}}

\DeclarePairedDelimiter\abs{|}{|}

\newcommand{\eto}[1]{\ensuremath{\mathrm{e}^{#1}}}

\newcommand{\threeD}{3D}
\newcommand{\twoPoneD}{(2+1)D}

\makeatletter%
\makeatletter%
\begin{document}

\addtolength{\baselineskip}{0.20\baselineskip}

\hfill \today

\begin{center}

\vspace{24pt}

{\LARGE {\bf Lattice Field Theory Analysis of the Chiral Heisenberg Model} }

\end{center}

\vspace{18pt}

\centerline{{\bf Simon Hands$^{*}$, Johann Ostmeyer$^{\dagger}$}}

\vspace{20pt}

\centerline{$^{*}$Department of Mathematical Sciences, University of Liverpool,}
\centerline{Liverpool L69 3BX, U.K.}

\centerline{$^{\dagger}$Helmholtz-Institut f\"{u}r Strahlen- und Kernphysik,}
\centerline{University of Bonn, 53115 Bonn, Germany}

\begin{center}  
{\bf Abstract}
\end{center}

\noindent
Motivated by ongoing interest in the universal behaviour of the Hubbard model of spinning electrons on
honeycomb and $\pi$-flux lattices at the semi-metal -- Mott insulator phase
transition, 
we formulate the \threeD~chiral Heisenberg model, a theory of relativistic
fermions in three spacetime dimensions,
as a lattice field theory using  domain wall fermions. 
The contact interaction term preserves an SU(2) global symmetry. 
We perform numerical
simulations using the Rational Hybrid Monte Carlo algorithm on system sizes
$L^3\times L_s$ with $L\in\{8,\ldots,24\}$ and domain wall separation
$L_s\in\{8,16,24\}$.
We locate the phase transition corresponding to spontaneous SU(2)$\to$U(1)
breaking, yielding critical exponent estimates $\nu^{-1}=0.63(3)$,
$\eta_\Phi=1.42(8)$. These values are considerably removed from estimates
obtained from simulations performed in \twoPoneD, ie.\ with the time and
spatial directions
treated differently, but align more closely with analytic estimates
obtained using \threeD\ covariant field theory. We also present first results for the fermion
correlator, ultimately needed for the determination of the exponent $\eta_\Psi$,
highlighting the need to rotate the fermion source to a common reference
direction in isospace in order to obtain a signal.

\bigskip


\vfill

\newpage

\section{Introduction}
The Hubbard model is a key tool in the modelling of
strongly-correlated electronic
structure in materials. At half filling the Hamiltonian is of the form
\begin{equation}
H=-\sum_{\langle
i,j\rangle;\sigma=\uparrow,\downarrow}t_{ij}\left[c^\dagger_{i\sigma}c_{j\sigma}+{\rm
h.c.}\right]
+U\sum_i
(n_{i\uparrow}-{\textstyle{1\over2}})(n_{i\downarrow}-{\textstyle{1\over2}}),
\label{eq:Hubbard}
\end{equation}
where $c^\dagger,c$ create and annihilate electrons of spin $\sigma$ at the
sites $i$ of some lattice, the tunnelling $t_{ij}$ is non-zero only on
nearest neighbour sites $\langle i,j\rangle$, and the number of electrons at
site $i$ is $n_{i\sigma}=c^\dagger_{i\sigma}c_{i\sigma}\in\{0,1\}$.
The form of the on-site two-body interaction proportional to $U$ is constrained by
the Pauli exclusion principle. In two-dimensional systems there is particular
interest in the cases where either $i,j$ denote the sites of a honeycomb
lattice with constant $t_{ij}=t$, or $i,j$ live on a square lattice and the product of the
four $t_{ij}$ around any elementary plaquette is constrained to equal -1,
yielding a ``$\pi$-flux''. In either case the resulting low-energy electron
dispersion vanishes at two {\em Dirac points\/} in the first Brillouin zone,
around which the dispersion is linear, and propagation of single-particle excitations is described
by the Dirac equation in the long-wavelength limit.
At half-filling the Fermi energy equals zero
and particle or hole excitations can be interpreted as particles/antiparticles
in a relativistic quantum field theory. For weak interactions the resulting
system is a semi-metal. As $U$ is increased, there may be a transition to
a Mott insulator phase triggered by the opening of a gap at the Dirac points.

The field theory underlying the universality class of the phase transition is
determined by the number of low-energy electronic degrees of freedom. For each
of the lattices mentioned above there are two sites per unit cell and two Dirac
points per Brillouin zone; finally there are two spin components, yielding
$2\times2\times2$ = 8 components. In $d=2+1$ dimensions this naturally leads to a
description in terms of $N=2$ flavors of four-component Dirac spinors, to be set
out in Sec.~\ref{sec:formulation} below.  However, it is the symmetry of the
interaction and its pattern of gap-driven breaking which
characterise the universality class in this case. Defining the spin at site $i$,
\begin{equation}
\vec S_i={1\over2}c^\dagger_{i\alpha}\vec\tau_{\alpha\beta}c_{i\beta}, \end{equation}
where $\vec\tau$ are Pauli matrices, we find, using 
$\sum_a\tau^a_{\alpha\beta}\tau^a_{\gamma\delta}
\equiv2\delta_{\alpha\delta}\delta_{\beta\gamma}-\delta_{\alpha\beta}\delta_{\gamma\delta}$,
that up to an irrelevant constant the interaction in (\ref{eq:Hubbard}) can be
recast as $-{2U\over3}\sum_i\vec S_i\cdot\vec S_i$. Classically this resembles the 
Hamiltonian of the Heisenberg model; the corresponding  effective field theory
of interacting relativistic fermions
is known as the {\em chiral Heisenberg model\/}.

There have been many attempts to characterise the chiral Heisenberg universality
class by estimating critical exponents. It's conventional to focus on the
correlation length exponent $\nu$, and the exponent $\eta_\Phi$ defined by
critical correlations of the order parameter field $\Phi$:
\begin{equation}
\langle\Phi(0)\Phi(r)\rangle_{U=U_c}\propto r^{-(d-2+\eta_\Phi)}\sim
r^{-(1+\eta_\Phi)}.
\end{equation}
Another exponent $\eta_\Psi$ characterising critical correlations of the
fermion field is needed for complete specification; since our results for the
fermion correlator reviewed in Sec.~\ref{sec:fermions} below are exploratory we will
not discuss $\eta_\Psi$ much in this work. 

One group of estimates, motivated by
closeness to the microscopic physics, emerges from
numerical simulation of the Hubbard model (\ref{eq:Hubbard}) using either
quantum Monte Carlo (QMC), e.g.\
\cite{ParisenToldin:2014nkk,Otsuka:2015iba,Liu:2018sww,Otsuka:2020lhc,Xu:2020qbj,Liu:2021npk,Wang:2026jwf},
or hybrid Monte Carlo (HMC) simulations~\cite{Ostmeyer:2020uov,Ostmeyer:2021efs,Buividovich:2018yar,Buividovich:2018crq}
employing imaginary time.
A distinct group of QMC simulations~\cite{Lang:2018csk,Lang:2025cwl} employs the
so-called SLAC formulation of the fermion hopping term, which better captures
the symmetries of relativistic fermions at the price of introducing
non-local tunnelling.
The Mott insulator phase is manifested in studies of solid state materials
like layered
WSe${}_2$~\cite{Biedermann:2025dma,hawashin2025relativisticmotttransitionhighorder}
and has recently even been observed experimentally~\cite{Ma2025RelativisticMT},
though critical exponents could not be extracted to date.  A common feature of
all these \twoPoneD~simulations is that since ultimately a time continuum limit
is needed, distinct approaches are taken to discretising time and space
directions.  As reviewed in Sec.~\ref{sec:comparison} and
Fig.~\ref{fig:compendium} below, these simulation-based estimates tend to yield
exponent estimates $\nu^{-1}\gtrsim0.8$, $\eta_\Phi\lesssim1$, with a large
spread, which has been attributed to the prevalence of severe finite volume
effects~\cite{Wang:2026jwf}.

Another group of estimates emerges from analytic or semianalytic calculations 
rooted in Euclidean quantum field theory: these include loop expansions about
renormalisable models in the Gross-Neveu (GN) class in $d=2$~\cite{Ladovrechis:2022aof}, 
or Gross-Neveu-Yukawa (GNY) class\footnote{An exhaustive classification of GNY
models setting out their field content and interaction symmetries can be found in 
\cite{Mitchell:2025oao}.} in $d=4$~\cite{Zerf:2017zqi,Ladovrechis:2022aof},
large-$N$ expansions in fixed dimensionality $d$~\cite{Gracey:2018qba}, or
functional renormalisation group (FRG)
approaches~\cite{Janssen:2014gea,Knorr:2017yze}. A unifying feature of such
\threeD ~approaches is
that Euclidean time is treated on the same footing as the spatial dimensions, up
to boundary conditions.  Since we're mainly interested in critical properties, the
continuum limit may be taken uniformly across both temporal and spatial directions.
Fig.~\ref{fig:compendium} shows that field
theory-based estimates tend to have $\nu^{-1}\lesssim0.9$, $\eta_\Phi\gtrsim1$,
again with a large spread.

In this paper we will furnish a further estimate, based on a \threeD~lattice field theory
simulation of the appropriate Euclidean theory. This aligns us very much
with the latter group, but this time based on a method with a completely
different set of systematic but theoretically controllable errors, dictated by
available computational resource. Our approach treats fermions
$\psi,\bar\psi$ as independent Grassmann fields, formulated using {\rm Domain
Wall Fermions\/} (DWF), a lattice-based discretisation 
specifically devised to reproduce the symmetries of continuum
Dirac fermions as closely as possible. 
The DWF are defined on a $L^3\times L_s$ 
hypercubic lattice, where open boundary conditions along an artificial third
direction
result in two domain walls separated by $L_s$. The dynamics of the
target theory are then carried by near-zero modes of the DWF operator
localised on either wall. It has been demonstrated both
numerically~\cite{Hands:2015qha}
and analytically~\cite{Hands:2015dyp} that in the limit $L_s \to\infty$ the desired
U($2N$) symmetry
of weakly-interacting relativistic Dirac fermions is restored, in the sense that
the resulting effective 3D Dirac operator $D$ respects a three-dimensional
generalisation of the Ginsparg-Wilson relations~\cite{Ginsparg_Wilson}.
DWF have previously been applied to the study of critical behaviour in the 3D GN model~\cite{Vranas:1999nx,
Hands:2016foa} and the 3D Thirring
model~\cite{Hands:2018vrd,Hands:2020itv,Hands:2025aje}.

In what follows, in Sec.~\ref{sec:formulation} we first specify the Euclidean action
of the continuum \threeD\
field theory in the chiral Heisenberg universality class, sharing the same
symmetries and particle content at low energies as the original Hubbard model (\ref{eq:Hubbard}),
followed by its DWF discretisation. The action is phrased in terms of a real scalar 
auxiliary boson triplet $\vec\phi$, which using particle physics parlance  we will refer to as an {\em
isotriplet\/}. The numerical simulation is performed using the
rational hybrid Monte Carlo (RHMC) algorithm~\cite{Kennedy:1998cu}; some
simplifying features applicable since  $\vec\phi$ is only defined
on the domain walls are noted. Appendix~\ref{app:A} shows that the continuum
model has a real positive fermion determinant, and that the Sign Problem
introduced by DWF is likely to be mild and safe to ignore. As already noted, the approach differs from
most simulations in condensed matter physics in treating spatial and
temporal directions alike; it also differs from particle physics applications in
having no symmetry breaking parameter, such as an explicit fermion mass, to
guide the direction of symmetry breaking. In Sec.~\ref{sec:FVS} it is shown how
in the resulting simulation the auxiliary $\vec\phi$ explores the manifold of allowed ground states in a
fairly uniform manner, so that a pragmatic choice of pseudo-order parameter
$\vert\Phi\vert$ is needed, defined in (\ref{eq:pseudo}) below. A series of
simulations of varying system size $L^3$, with $L\in\{8,\ldots,24\}$,  and domain
wall separation $L_s\in\{8,16,24\}$ are
performed with $\vert\Phi\vert$ as the principle observable, and a critical
coupling is identified, first using the  Binder cumulant and next with
more precision using a finite volume scaling analysis, leading to estimates for
critical  exponents. Particular attention is paid to the
stability of the results as $L_s$ is increased towards the limit where the
symmetries of the continuum model are expected to be restored. Next in
Sec.~\ref{sec:comparison} we set things in context by comparing our predictions
for $\nu^{-1}$ and $\eta_\Phi$  with a
compendium of other results found in the recent literature. 

In Sec.~\ref{sec:fermions} we switch attention to the fermion sector where ultimately
a controlled analysis will be needed to extract the exponent $\eta_\Psi$; a
possible route is sketched in \cite{Hands:2022fhq}. Our lattice sizes are too
small to reach the Euclidean time separations needed to establish critical
scaling; rather we present exploratory work,
first in Sec.~\ref{sec:fermioncorrelator} 
defining the correlation function to be studied
in terms of the component remaining after
tracing over Dirac indices, yielding a timeslice correlator $S_m(x_0)$ which is
anti-symmetric about the lattice midpoint $x_0=L/2$. In order to have a non-vanishing
signal following the drift of $\vec\phi$ around the vacuum manifold we have to introduce 
a rotation $S_m\mapsto\tilde S_m$ to align the source along the 3rd
isodirection. Further details are presented in Appendix~\ref{app:B}.
Finally Sec.~\ref{sec:THC} presents an 
analysis of $\tilde S_m$ for various couplings and lattice volumes
using the Truncated Hankel Correlator (THC)
method~\cite{Ostmeyer:2025igc}, and masses of the lowest excited
state and overlap amplitudes for the lowest excited state extracted.
Sec.~\ref{sec:summary} gives a brief summary of our conclusions.

\section{Formulation and Symmetry}
\label{sec:formulation}

Our starting point is the continuum Lagrangian density for the chiral Heisenberg
model:
\begin{equation}
{\cal
L}=\bar\psi(\partial_\mu\gamma_\mu\otimes\One_{2\times2}+g\vec\phi\cdot \One_{4\times4}\otimes\vec\tau)\psi+{1\over2}\vec\phi\cdot \vec\phi
\equiv\bar\psi {\cal M}\psi+{1\over2}\vec\phi\cdot \vec\phi.
\label{eq:Lcont}
\end{equation}
Here $\psi,\bar\psi$ represent two flavors of four-component spinor. 
In our notation $\gamma_A$ are 4$\times4$ traceless hermitian matrices
acting on spinor indices 
satisfying $\{\gamma_A,\gamma_B\}=\delta_{AB}$, with $A\in\{0,1,2,3,5\}$,
and the $\vec\tau$ act on flavor.
Spacetime indices are denoted by $\mu\in\{0,1,2\}$.
The following invariances of the kinetic term (ie. with $g=0$ in
(\ref{eq:Lcont})) generate a global U(2) invariance of the free fermion
Lagrangian:
\begin{eqnarray}
\psi\mapsto e^{i\alpha}\psi;\;\bar\psi\mapsto\bar\psi
e^{-i\alpha};\;\;&\phantom{=}&\;\;
\psi\mapsto e^{\alpha\gamma_3\gamma_5}\psi;\;\bar\psi\mapsto\bar\psi
e^{-\alpha\gamma_3\gamma_5};\nonumber\cr
\psi\mapsto e^{i\alpha\gamma_3}\psi;\;\bar\psi\mapsto\bar\psi
e^{i\alpha\gamma_3};\;\;&\phantom{=}&\;\;
\psi\mapsto e^{i\alpha\gamma_5}\psi;\;\bar\psi\mapsto\bar\psi
e^{i\alpha\gamma_5}.
\end{eqnarray}

In the interaction, $\vec\phi$ is a real 3-component scalar auxiliary field.
It is readily integrated out to recover a four-fermi contact interaction
resembling that of the Hubbard model (\ref{eq:Hubbard}), 
but the form (\ref{eq:Lcont}) is more
convenient to handle, with a manifest invariance 
under a global SU(2) rotation:
\begin{equation}
\psi\mapsto U\psi;\;\;\bar\psi\mapsto\bar\psi U^\dagger;\;\;\phi\mapsto O\phi
\label{eq:SU(2)}
\end{equation}
with $U\in{\rm SU}(2)$, $O_{ij}={1\over2}{\rm tr}[\tau_iU^\dagger\tau_jU]\in{\rm
SO}(3)$. This model corresponds to the case $N_D=1$, $M=3$ in the classification
of Gross-Neveu-Yukawa models presented in \cite{Mitchell:2025oao}.
Because $\phi$ is real the interaction term breaks the U(2) invariance of free
Dirac fermions in \threeD\ to U(1)$\otimes$U(1). 

Our simulations are performed using Domain Wall Fermions (DWF) on a $L^3\times L_s$
lattice, where open boundary conditions along an artificial third direction result in two
domain walls separated by $L_s$. 
In the limit
$L_s\to\infty$ the desired U($2$) symmetry of weakly-interacting relativistic
Dirac fermions is restored, in the sense that the resulting
effective \threeD\ Dirac
operator $D$ respects the three-dimensional generalisation of the
Ginsparg-Wilson relatons~\cite{Hands:2015dyp}:
\begin{equation}
\{\gamma_3,D\}=2aD\gamma_3D;\;\;\;
\{\gamma_5,D\}=2aD\gamma_5D;\;\;\;
[\gamma_3\gamma_5,D]=0.
\label{eq:GW}
\end{equation}
In the long-wavelength limit $aD\to0$ (where $a$ is the physical lattice
spacing, implicitly set to unity almost everywhere) this guarantees U($2$) restoration.

Here we present the discretised action $S=S_{\rm kin}+S_{\rm
int}+S_\phi=\bar\Psi{\cal M}\Psi+S_\phi$ 
in detail, following the discussion of 
\cite{Hands:2016foa}. 
The fermion kinetic term uses the
\threeD\ domain wall operator defined in~\cite{Hands:2015qha, Hands:2015dyp}:
\begin{equation}
S_{\rm kin}=
\sum_{x,y}\sum_{s,s^\prime}\bar\Psi(x,s)[\delta_{s,s^\prime}D_W(x\vert
y)+\delta_{x,y}D_3(s\vert
s^\prime)]\Psi(y,s^\prime),
\label{eq:SDWF}
\end{equation}
where 
$\Psi,\bar\Psi$ are four-component isodoublet spinors defined in 2+1+1
dimensions. The relation with the fields $\psi,\bar\psi$ living in the \twoPoneD\
target space is
\begin{equation}
\psi(x)={\cal P}_-\Psi(x,1)+{\cal P}_+\Psi(x,L_s);\;\;
\bar\psi(x)=\bar\Psi(x,L_s){\cal P}_-+\bar\Psi(x,1){\cal P}_+,
\label{eq:projection}
\end{equation}
with projectors ${\cal P}_\pm={1\over2}(1\pm\gamma_3)$.
$D_W$ is the \threeD\ Wilson operator defined on spacetime volume
$V$
\begin{equation}
D_W(M)_{x,y}=-{1\over2}\sum_\mu
\left[(1-\gamma_\mu)\delta_{x+\hat\mu,y}+(1+\gamma_\mu)\delta_{x-\hat\mu,y}
\right]
+(3-M)\delta_{x,y},
\label{eq:Ddw}
\end{equation}
with the domain wall height parameter $M$ set to unity throughout,
and $D_3$ controls hopping along the third dimension between the walls at
$s=1$ and $s=L_s$:
\begin{equation}
D_{3\,s,s^\prime}
=-\left[{\cal P}_-\delta_{s+1,s^\prime}
(1-\delta_{s^\prime,L_s})
+{\cal P}_+\delta_{s-1,s^\prime}(1-\delta_{s^\prime,1})\right]
+\delta_{s,s^\prime}.
\label{eq:D3dw}
\end{equation}
The DWF formulation of the interaction term 
is defined solely in terms of fields on the domain walls,
\begin{equation}
S_{\rm
int}=\sum_x\vec\phi(x)\cdot [\bar\Psi(x,L_s){\cal
P}_-\vec\tau\Psi(x,1)+\bar\Psi(x,1){\cal P}_+\vec\tau\Psi(x,L_s)].
\label{eq:SGNint} 
\end{equation}
As originally noted in \cite{Vranas:1999nx}, 
restriction of the scalar degrees of freedom to the domain walls implies that the Pauli-Villars
kernel normally needed to cancel unphysical contributions from bulk modes has  
no $\phi$-dependence, and therefore can be omitted from the simulation
dynamics. Finally, the auxiliary action 
\begin{equation}
S_\phi=\beta\sum_x\vec\phi(x)\cdot \vec\phi(x).
\end{equation}
with the dimensionless parameter $\beta=ag^{-2}$.
All terms in $S$ are invariant under the SU(2) global symmetry (\ref{eq:SU(2)}).

It is shown in Appendix~\ref{app:A} that the continuum theory (\ref{eq:Lcont}) is free of a sign
problem, while the DWF-regularised version has a mild sign problem 
expected to vanish as $L_s\to\infty$. Accordingly we simulate $N=1$ flavor (ie.
a single isodoublet of fermion 4-spinors)
using the Rational Hybrid Monte Carlo (RHMC)~\cite{Kennedy:1998cu} 
algorithm which updates using the positive functional measure $({\rm
det}{\cal M}^\dagger{\cal M})^{1\over2}$. Further details of the algorithm are presented in
\cite{Hands:2018vrd}.

\section{Numerical Results}
\subsection{Finite Volume Scaling}
\label{sec:FVS}
\begin{figure}[t] 
\centering
\subfigure[$\beta=0.25$]
{\includegraphics[width=.45\textwidth]{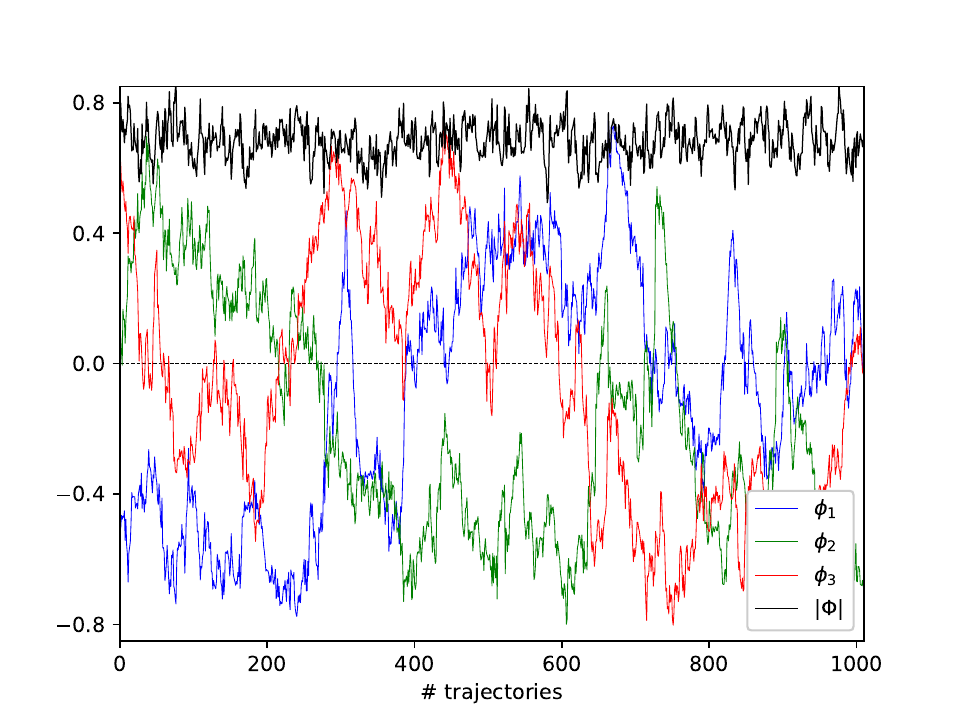}}
\hspace*{4pt}
\subfigure[$\beta=0.45$]
{\includegraphics[width=.45\textwidth]{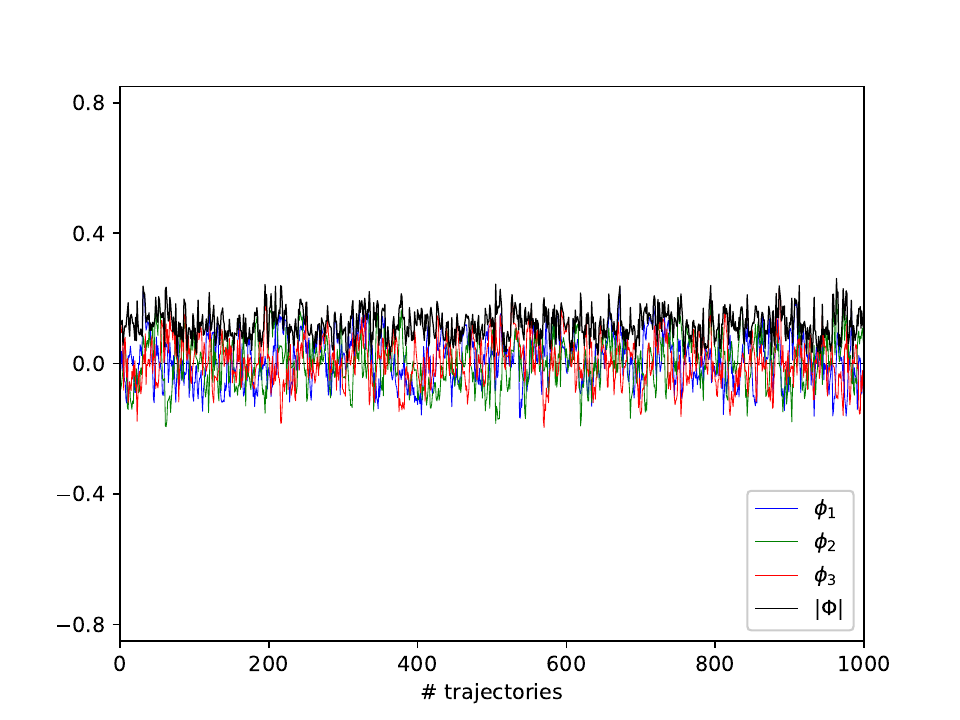}}
\caption{Scalar field histories over 1000 RHMC trajectories
on $12^3\times8$. Coloured lines show the component
$\phi_{i=1\ldots3}$;
the black line is $\vert\Phi\vert$ (see eq.~\eqref{eq:pseudo}).}
\label{fig:drift} 
\end{figure}

Our goal is to study the spontaneous breaking SU(2)$\to$U(1) expected
at some critical coupling $\beta_c$. In the infinite volume limit the symmetry
broken phase is signalled by long-range order, namely a non-zero scalar field
expectation $\langle\vec\phi\rangle\not=\vec0$. In a finite system, however, in
the absence of explicit symmetry breaking such as a fermion mass, the mean 
direction $\vec\phi/\vert\phi\vert$ drifts around the SO(3) manifold, so on long
enough
simulation timescales $\langle\vec\phi\rangle\approx0$, as illustrated in
Fig.~\ref{fig:drift}. 
This demonstrates that our simulations are ergodic and cover the entire phase space efficiently.
To study symmetry
breaking a pragmatic alternative is therefore the {\em pseudo-order parameter\/}
\begin{equation}
\vert\Phi\vert
={1\over L^3}\sqrt{\sum_{a=1}^3\left(\sum_x\phi_a(x)\right)^2}.
\label{eq:pseudo}
\end{equation}
\begin{figure}[t] 
\centering
\includegraphics[width=.9\textwidth]{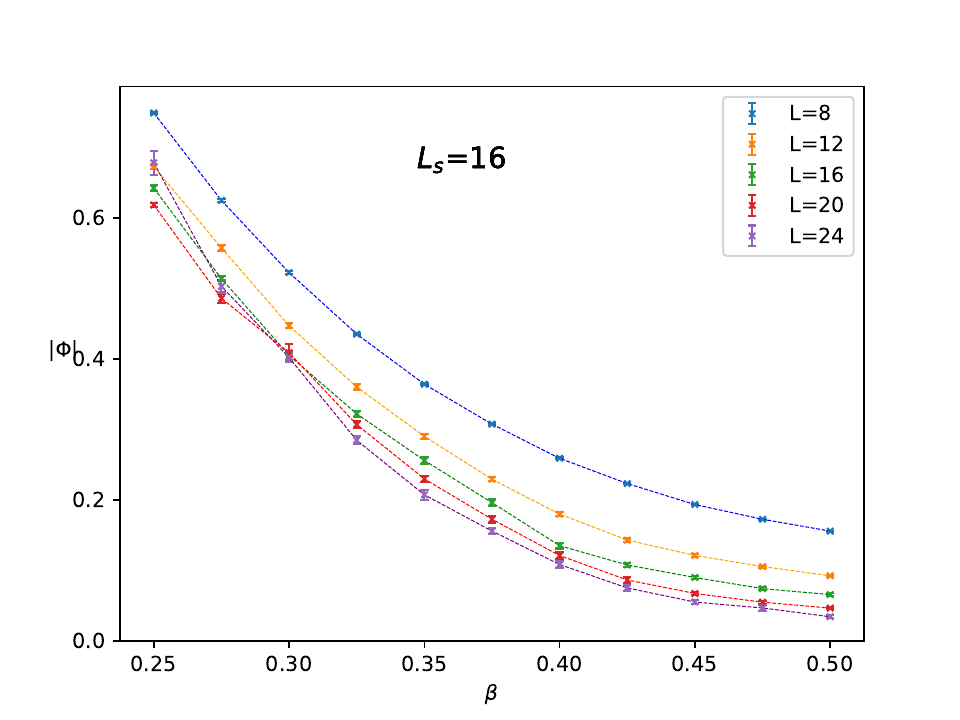} 
\caption{Pseudo-order parameter $\vert\Phi\vert$ (see eq.~\eqref{eq:pseudo}) vs.\ coupling $\beta$ for 5 spacetime volumes $L^3$ with
	$L_s=16$. Data from $L_s=8,24$ are virtually indistinguishable.}
\label{fig:dataset_16} 
\end{figure}

Pilot studies of the \threeD\ Gross-Neveu model using DWF~\cite{Hands:2016foa} 
have found the large-$L_s$ limit needed for the GW symmetries
(\ref{eq:GW}) is for practical purposes recovered by $L_s\sim O(20)$, which is
relatively modest compared with recent studies of the \threeD\ Thirring
model~\cite{Hands:2025aje}\footnote{The $L_s\to\infty$ limit in GN and Thirring
models is best achieved using bilinear order parameters of the form $\bar\psi
i\gamma_3\psi$. Here the scalar density
$\vec\phi\propto\bar\psi(\One\otimes\vec\tau)\psi$ is just as effective; the essential
requirement is the tracelessness of the kernel~\cite{Hands:2015dyp}.}. 
To corroborate this we make separate simulations
using $L_s=8,16$ and 24.

We have simulated on $L^3\times L_s$ 
using $L=8,12,16,20,24$, with $\vert\Phi\vert$ the principal
observable. For each of $L_s=8,16,24$ a total of 55 datasets were generated,
each of between $\num{300}$ (for $L=L_s=24$) and $\num{50000}$ (for $L=L_s=8$) measurements.
The $L_s=16$ dataset is shown in
Fig.~\ref{fig:dataset_16}.

\begin{figure}[t] 
\centering
\begingroup
  \inputencoding{latin1}%
  \makeatletter
  \providecommand\color[2][]{%
    \GenericError{(gnuplot) \space\space\space\@spaces}{%
      Package color not loaded in conjunction with
      terminal option `colourtext'%
    }{See the gnuplot documentation for explanation.%
    }{Either use 'blacktext' in gnuplot or load the package
      color.sty in LaTeX.}%
    \renewcommand\color[2][]{}%
  }%
  \providecommand\includegraphics[2][]{%
    \GenericError{(gnuplot) \space\space\space\@spaces}{%
      Package graphicx or graphics not loaded%
    }{See the gnuplot documentation for explanation.%
    }{The gnuplot epslatex terminal needs graphicx.sty or graphics.sty.}%
    \renewcommand\includegraphics[2][]{}%
  }%
  \providecommand\rotatebox[2]{#2}%
  \@ifundefined{ifGPcolor}{%
    \newif\ifGPcolor
    \GPcolortrue
  }{}%
  \@ifundefined{ifGPblacktext}{%
    \newif\ifGPblacktext
    \GPblacktexttrue
  }{}%
  \let\gplgaddtomacro\g@addto@macro
  \gdef\gplbacktext{}%
  \gdef\gplfronttext{}%
  \makeatother
  \ifGPblacktext
    \def\colorrgb#1{}%
    \def\colorgray#1{}%
  \else
    \ifGPcolor
      \def\colorrgb#1{\color[rgb]{#1}}%
      \def\colorgray#1{\color[gray]{#1}}%
      \expandafter\def\csname LTw\endcsname{\color{white}}%
      \expandafter\def\csname LTb\endcsname{\color{black}}%
      \expandafter\def\csname LTa\endcsname{\color{black}}%
      \expandafter\def\csname LT0\endcsname{\color[rgb]{1,0,0}}%
      \expandafter\def\csname LT1\endcsname{\color[rgb]{0,1,0}}%
      \expandafter\def\csname LT2\endcsname{\color[rgb]{0,0,1}}%
      \expandafter\def\csname LT3\endcsname{\color[rgb]{1,0,1}}%
      \expandafter\def\csname LT4\endcsname{\color[rgb]{0,1,1}}%
      \expandafter\def\csname LT5\endcsname{\color[rgb]{1,1,0}}%
      \expandafter\def\csname LT6\endcsname{\color[rgb]{0,0,0}}%
      \expandafter\def\csname LT7\endcsname{\color[rgb]{1,0.3,0}}%
      \expandafter\def\csname LT8\endcsname{\color[rgb]{0.5,0.5,0.5}}%
    \else
      \def\colorrgb#1{\color{black}}%
      \def\colorgray#1{\color[gray]{#1}}%
      \expandafter\def\csname LTw\endcsname{\color{white}}%
      \expandafter\def\csname LTb\endcsname{\color{black}}%
      \expandafter\def\csname LTa\endcsname{\color{black}}%
      \expandafter\def\csname LT0\endcsname{\color{black}}%
      \expandafter\def\csname LT1\endcsname{\color{black}}%
      \expandafter\def\csname LT2\endcsname{\color{black}}%
      \expandafter\def\csname LT3\endcsname{\color{black}}%
      \expandafter\def\csname LT4\endcsname{\color{black}}%
      \expandafter\def\csname LT5\endcsname{\color{black}}%
      \expandafter\def\csname LT6\endcsname{\color{black}}%
      \expandafter\def\csname LT7\endcsname{\color{black}}%
      \expandafter\def\csname LT8\endcsname{\color{black}}%
    \fi
  \fi
    \setlength{\unitlength}{0.0500bp}%
    \ifx\gptboxheight\undefined%
      \newlength{\gptboxheight}%
      \newlength{\gptboxwidth}%
      \newsavebox{\gptboxtext}%
    \fi%
    \setlength{\fboxrule}{0.5pt}%
    \setlength{\fboxsep}{1pt}%
    \definecolor{tbcol}{rgb}{1,1,1}%
\begin{picture}(7200.00,5040.00)%
    \gplgaddtomacro\gplbacktext{%
      \csname LTb\endcsname%
      \put(946,704){\makebox(0,0)[r]{\strut{}$0.4$}}%
      \csname LTb\endcsname%
      \put(946,1390){\makebox(0,0)[r]{\strut{}$0.45$}}%
      \csname LTb\endcsname%
      \put(946,2076){\makebox(0,0)[r]{\strut{}$0.5$}}%
      \csname LTb\endcsname%
      \put(946,2762){\makebox(0,0)[r]{\strut{}$0.55$}}%
      \csname LTb\endcsname%
      \put(946,3447){\makebox(0,0)[r]{\strut{}$0.6$}}%
      \csname LTb\endcsname%
      \put(946,4133){\makebox(0,0)[r]{\strut{}$0.65$}}%
      \csname LTb\endcsname%
      \put(946,4819){\makebox(0,0)[r]{\strut{}$0.7$}}%
      \csname LTb\endcsname%
      \put(1078,484){\makebox(0,0){\strut{}$0.25$}}%
      \csname LTb\endcsname%
      \put(2223,484){\makebox(0,0){\strut{}$0.3$}}%
      \csname LTb\endcsname%
      \put(3368,484){\makebox(0,0){\strut{}$0.35$}}%
      \csname LTb\endcsname%
      \put(4513,484){\makebox(0,0){\strut{}$0.4$}}%
      \csname LTb\endcsname%
      \put(5658,484){\makebox(0,0){\strut{}$0.45$}}%
      \csname LTb\endcsname%
      \put(6803,484){\makebox(0,0){\strut{}$0.5$}}%
    }%
    \gplgaddtomacro\gplfronttext{%
      \csname LTb\endcsname%
      \put(2002,1922){\makebox(0,0)[r]{\strut{}  $L=8$}}%
      \csname LTb\endcsname%
      \put(2002,1592){\makebox(0,0)[r]{\strut{}  $L=12$}}%
      \csname LTb\endcsname%
      \put(2002,1262){\makebox(0,0)[r]{\strut{}  $L=16$}}%
      \csname LTb\endcsname%
      \put(2002,932){\makebox(0,0)[r]{\strut{}  $L=20$}}%
      \csname LTb\endcsname%
      \put(209,2761){\rotatebox{-270.00}{\makebox(0,0){\strut{}$B$}}}%
      \put(3940,154){\makebox(0,0){\strut{}$\beta$}}%
    }%
    \gplbacktext
    \put(0,0){\includegraphics[width={360.00bp},height={252.00bp}]{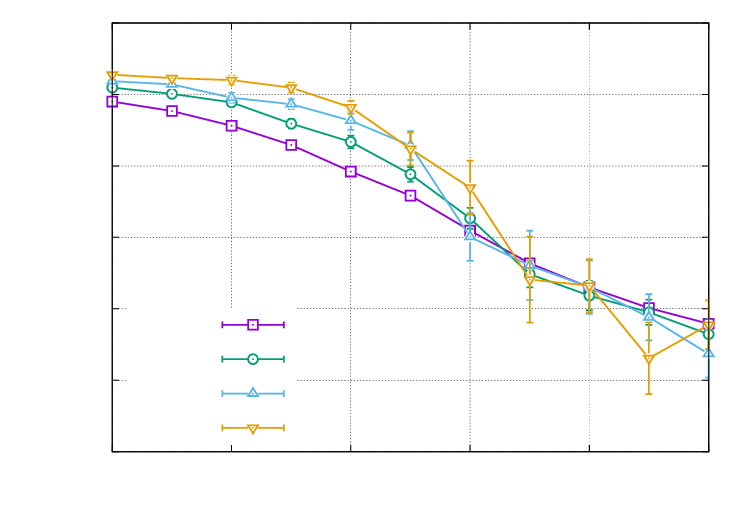}}%
    \gplfronttext
  \end{picture}%
\endgroup
 \caption{Binder cumulant $B$ as a function of the coupling $\beta$ for $L_s=16$.
	The plots look very similar for $L_s=8,24$.
	Error bars for $L\le 12$ are too small to be visible on this scale, results for $L=24$ are very noisy and have been omitted.}
\label{fig:binder} 
\end{figure}

First consider the Binder cumulant
\begin{align}
	B &= 1 - \frac{\erwartung{\abs*{\Phi}^4}}{3\erwartung{\abs*{\Phi}^2}^2}\,,
\end{align}
proportional to the excess kurtosis (the 4th moment of the distribution) of the order parameter.
At a phase transition, $B$ becomes volume-independent and, in consequence,
lines corresponding to different $L$ should cross at the critical coupling $\beta_c$.
The Binder cumulant has been plotted for $L_s=16$ in
Fig.~\ref{fig:binder}, which suggests a critical coupling
$\beta_c\approx0.45$.
The noise level of $B$ is rather high so that the resulting estimation of $\beta_c$ has a low precision.
It can, however, serve as an independent sanity check for the critical coupling extracted in the following via the far more accurate finite-volume scaling (FVS).

To extract critical exponents $\beta_\Phi$ and $\nu$, for each $L_s$ we fit to a
FVS form
\begin{equation} \vert\Phi\vert
L^{\beta_\Phi\over\nu}=f(tL^{1\over\nu}),\label{eq:FVS}
\end{equation}
with the dimensionless
reduced coupling $t\equiv(\beta-\beta_c)/\beta_c$.  We performed fits using both
quadratic $f(x)=A+Bx+Cx^2$ and cubic $f(x)=A+Bx+Cx^2+Dx^3$ {\em Ans\"atze\/}, in
all cases including the 36 datapoints from within a window $\vert
x\vert\leq1.5$.  To increase the fit stability, we used initial guesses for the
critical exponents obtained with the method from Ref.~\cite{Ostmeyer_2020},
which
approximates the data by cubic splines for every $L$ and minimises the integrals
between the different resulting curves.  These results were also compatible with
our subsequent fits which makes us confident that the polynomial approximations
did not introduce a significant bias.

\begin{figure}[hp] 
\centering 
\hfill
\includegraphics[width=.48\textwidth,trim=0 0 1.5cm 1cm, clip]{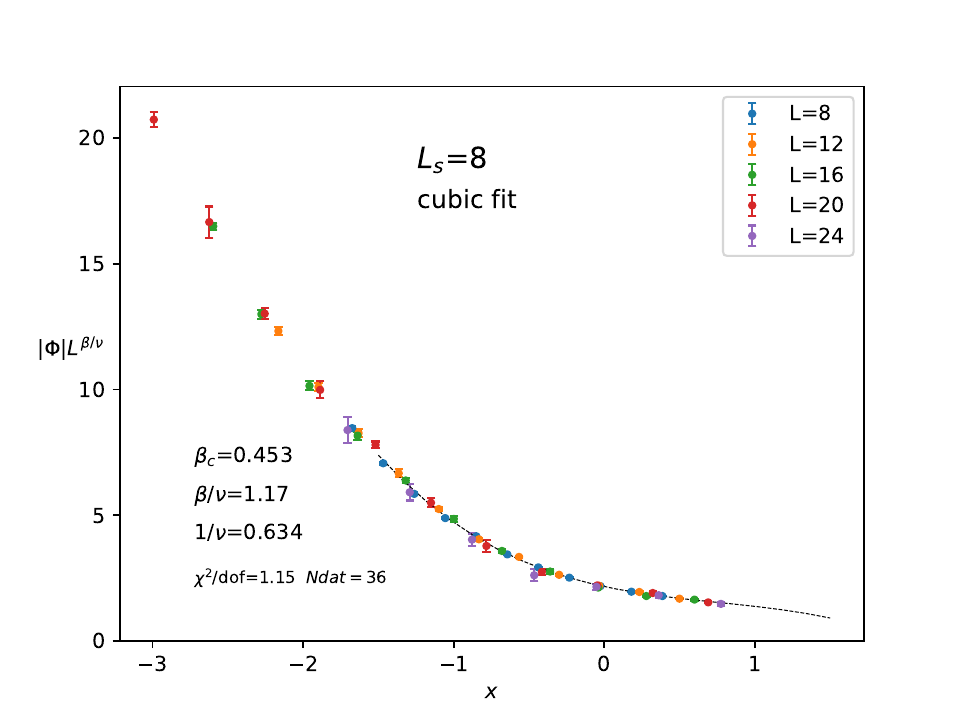}
\hfill
\includegraphics[width=.48\textwidth,trim=0 0 1.5cm 1cm, clip]{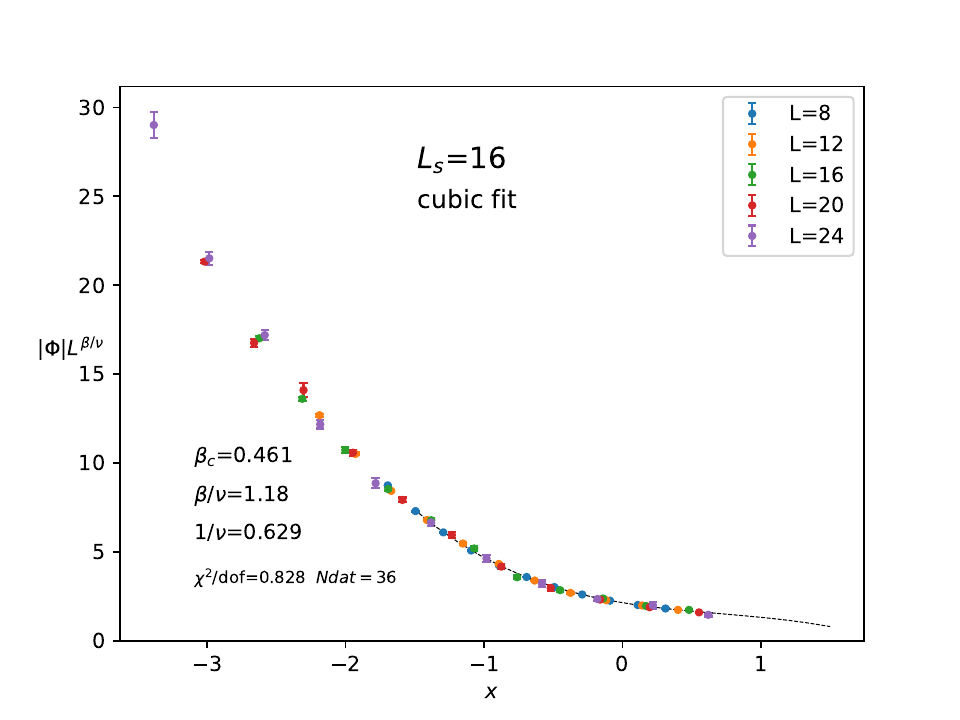} 
\hfill
\includegraphics[width=.96\textwidth,trim=0 0 1.5cm 1cm, clip]{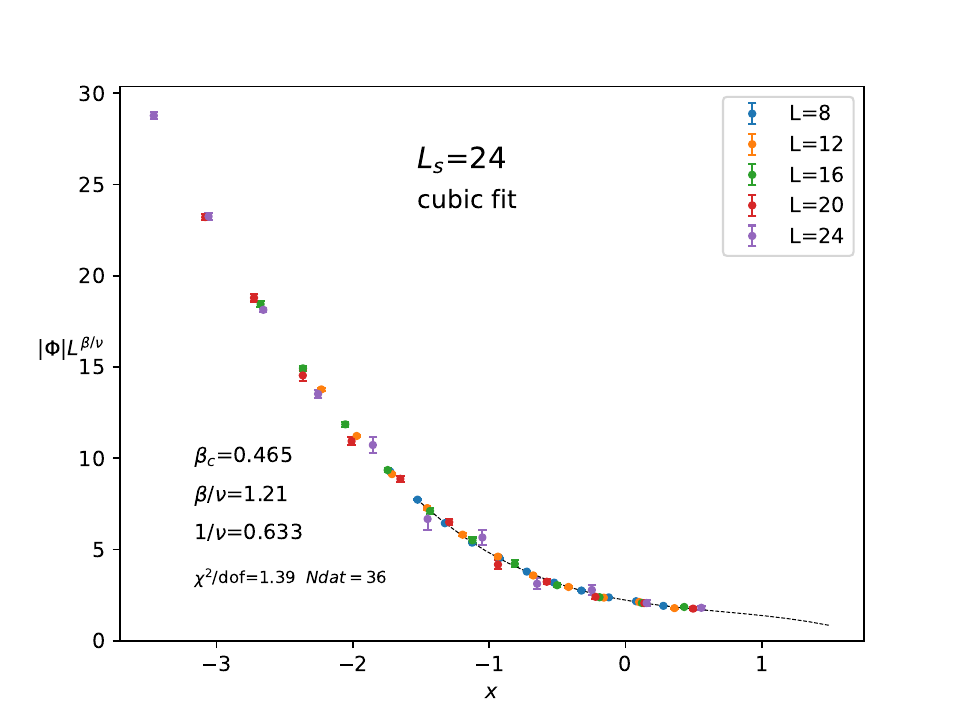}
\caption{Data collapse for the FVS~\eqref{eq:FVS} using cubic fits.
	Results are provided in Tab.~\ref{tab:fsscubic}.
	The panels differ in domain wall separation $L_s$.}
\label{fig:plot5_Ls24} 
\end{figure}
The fitted values for the critical
parameters are given in Tabs.~\ref{tab:fssquadratic},\ref{tab:fsscubic}, and
data collapse for the cubic fits plotted in
Fig.~\ref{fig:plot5_Ls24}. The $\chi^2$ values quoted for
the cubic fits were obtained first by fitting for all 7 parameters and
then fixing the critical parameters and refitting $A,B,C,D$, yielding values
O(10\%) smaller.
\begin{table}[t]
\begin{center}
{\begin{tabular}{@{}ccccc@{}} 
\hline
$L_s$  & $\beta_c$ & $\beta_\Phi/\nu$ & $1/\nu$ & $\chi^2/{\rm dof}$     \cr
 \hline
8 & 0.470(6) & 1.22(2) & 0.631(26) & 3.3 \cr 
16 & 0.468(5) & 1.20(2) & 0.627(22) & 5.6 \cr 
24 & 0.469(6) & 1.21(2) & 0.630(28) & 4.6 \cr 
\hline
\end{tabular} 
\caption{Critical coupling $\beta_c$ and exponents $\nu^{-1}$, $\eta_\Phi$ obtained via FVS~\eqref{eq:FVS} fits using a quadratic scaling function.}
\label{tab:fssquadratic}}
\end{center}
\begin{center}
{\begin{tabular}{@{}ccccc@{}} 
\hline
$L_s$  & $\beta_c$ & $\beta_\Phi/\nu$ & $1/\nu$ & $\chi^2/{\rm dof}$     \cr
 \hline
8 & 0.453(9) & 1.17(4) & 0.634(28) & 1.2 \cr 
16 & 0.461(5) & 1.18(4) & 0.629(22) & 0.8 \cr 
24 & 0.465(11) & 1.21(4) & 0.633(26) & 1.4 \cr 
\hline
\end{tabular} 
\caption{Critical coupling $\beta_c$ and exponents $\nu^{-1}$, $\eta_\Phi$ obtained via FVS~\eqref{eq:FVS} fits using a cubic scaling function.
Fits are visualised in Fig.~\ref{fig:plot5_Ls24}.}
\label{tab:fsscubic}}
\end{center}
\end{table}
The cubic fits are of significantly better quality than the quadratic, 
but the fitted critical parameters are compatible. Data has been taken in both
broken and symetric phases.
Importantly and encouragingly, there is no sign of any systematic drift
of the fitted critical parameters as $L_s$ is increased from 8 to 24, which 
suggests the results are obtained in a regime where the 
global symmetries are those of continuum Dirac fermions. This has proved to be a much
more troublesome issue in DWF simulations of the \threeD\ Thirring
model~\cite{Hands:2025aje}, where data from $L_s$ as large as 120 were required
for the analysis. We quote the $L_s=24$ results
\begin{equation}
\beta_c=0.465(11);\;\;\;{\beta_\Phi\over\nu}=1.21(4);\;\;\;{1\over\nu}=0.633(26)
\end{equation}
using  the errors output from the least squares routine.

\subsection{Comparison with Previous Work}
\label{sec:comparison}
\begin{table}[t]
\begin{center}
{\begin{tabular}{@{}llll@{}} 
\hline
Method & Year & $\nu^{-1}$ & $\eta_\Phi$ \cr
\hline
\hline
RHMC (DWF)~[this work] & 2026 & 0.63(3) & 1.42(8)\cr
$(2+\varepsilon),(4-\varepsilon)$ interpolate~\cite{Ladovrechis:2022aof} &
2022 & 0.83(12) & 1.01(6) \cr
$(4-\varepsilon)$ (4 loops)~\cite{Zerf:2017zqi} & 2017 & 0.64 & 0.98 \cr
Large-$N$ ($O(N^{-2})$ + Pad\'e)~\cite{Gracey:2018qba} & 2018 & 0.85 & 1.18 \cr
FRG (NLO derivative expansion)~\cite{Knorr:2017yze} & 2017 & 0.80 & 1.03 \cr
FRG (local potential approx)~\cite{Janssen:2014gea} & 2014 & 0.77 & 1.01 \cr
\hline
\hline
QMC (SLAC)~\cite{Lang:2025cwl} & 2025 & 0.98(2) & 0.73(1) \cr 
QMC (hcmb)~\cite{Wang:2026jwf} & 2026 & 0.90(10) & 0.79(2) \cr
QMC ($\pi$-flux)~\cite{Otsuka:2020lhc} & 2020 & 0.95(4) & 0.75(4)\cr
QMC (hcmb+$\pi$-flux)~\cite{Otsuka:2015iba} & 2015 & 0.98(1) & 0.47(7) \cr
QMC (hcmb)~\cite{Liu:2021npk} & 2021 & 1.11(4)  & 0.80(9) \cr
QMC (hcmb)~\cite{Liu:2018sww} & 2018 & 1.14(9) & 0.79(5) \cr
QMC (hcmb+$\pi$-flux)~\cite{ParisenToldin:2014nkk} & 2014 & 1.19(6) & 0.70(15) \cr
QMC ($\pi$-flux)~\cite{Xu:2020qbj} & 2020 & 1.10(8)  & 0.55(2) \cr
HMC (hcmb)~\cite{Ostmeyer:2021efs} & 2021 & 0.84(4) & 0.52(1) \cr
HMC (hcmb)~\cite{Ostmeyer:2020uov} & 2020 & 0.84(4) & 0.85(13) \cr
HMC (hcmb)~\cite{Buividovich:2018crq} & 2018 & 1.08 & 0.62 \cr
HMC (hcmb)~\cite{Buividovich:2018yar} & 2018 & 0.86 & 0.87(4) \cr
\hline
\end{tabular} 
\caption{Compendium of results for the critical exponents $\nu^{-1}$, $\eta_\Phi$ obtained
using various methods.}
\label{tab:compendium}}
\end{center}
\end{table}
\begin{figure}[t] 
\centering
\begingroup
  \inputencoding{latin1}%
  \makeatletter
  \providecommand\color[2][]{%
    \GenericError{(gnuplot) \space\space\space\@spaces}{%
      Package color not loaded in conjunction with
      terminal option `colourtext'%
    }{See the gnuplot documentation for explanation.%
    }{Either use 'blacktext' in gnuplot or load the package
      color.sty in LaTeX.}%
    \renewcommand\color[2][]{}%
  }%
  \providecommand\includegraphics[2][]{%
    \GenericError{(gnuplot) \space\space\space\@spaces}{%
      Package graphicx or graphics not loaded%
    }{See the gnuplot documentation for explanation.%
    }{The gnuplot epslatex terminal needs graphicx.sty or graphics.sty.}%
    \renewcommand\includegraphics[2][]{}%
  }%
  \providecommand\rotatebox[2]{#2}%
  \@ifundefined{ifGPcolor}{%
    \newif\ifGPcolor
    \GPcolortrue
  }{}%
  \@ifundefined{ifGPblacktext}{%
    \newif\ifGPblacktext
    \GPblacktexttrue
  }{}%
  \let\gplgaddtomacro\g@addto@macro
  \gdef\gplbacktext{}%
  \gdef\gplfronttext{}%
  \makeatother
  \ifGPblacktext
    \def\colorrgb#1{}%
    \def\colorgray#1{}%
  \else
    \ifGPcolor
      \def\colorrgb#1{\color[rgb]{#1}}%
      \def\colorgray#1{\color[gray]{#1}}%
      \expandafter\def\csname LTw\endcsname{\color{white}}%
      \expandafter\def\csname LTb\endcsname{\color{black}}%
      \expandafter\def\csname LTa\endcsname{\color{black}}%
      \expandafter\def\csname LT0\endcsname{\color[rgb]{1,0,0}}%
      \expandafter\def\csname LT1\endcsname{\color[rgb]{0,1,0}}%
      \expandafter\def\csname LT2\endcsname{\color[rgb]{0,0,1}}%
      \expandafter\def\csname LT3\endcsname{\color[rgb]{1,0,1}}%
      \expandafter\def\csname LT4\endcsname{\color[rgb]{0,1,1}}%
      \expandafter\def\csname LT5\endcsname{\color[rgb]{1,1,0}}%
      \expandafter\def\csname LT6\endcsname{\color[rgb]{0,0,0}}%
      \expandafter\def\csname LT7\endcsname{\color[rgb]{1,0.3,0}}%
      \expandafter\def\csname LT8\endcsname{\color[rgb]{0.5,0.5,0.5}}%
    \else
      \def\colorrgb#1{\color{black}}%
      \def\colorgray#1{\color[gray]{#1}}%
      \expandafter\def\csname LTw\endcsname{\color{white}}%
      \expandafter\def\csname LTb\endcsname{\color{black}}%
      \expandafter\def\csname LTa\endcsname{\color{black}}%
      \expandafter\def\csname LT0\endcsname{\color{black}}%
      \expandafter\def\csname LT1\endcsname{\color{black}}%
      \expandafter\def\csname LT2\endcsname{\color{black}}%
      \expandafter\def\csname LT3\endcsname{\color{black}}%
      \expandafter\def\csname LT4\endcsname{\color{black}}%
      \expandafter\def\csname LT5\endcsname{\color{black}}%
      \expandafter\def\csname LT6\endcsname{\color{black}}%
      \expandafter\def\csname LT7\endcsname{\color{black}}%
      \expandafter\def\csname LT8\endcsname{\color{black}}%
    \fi
  \fi
    \setlength{\unitlength}{0.0500bp}%
    \ifx\gptboxheight\undefined%
      \newlength{\gptboxheight}%
      \newlength{\gptboxwidth}%
      \newsavebox{\gptboxtext}%
    \fi%
    \setlength{\fboxrule}{0.5pt}%
    \setlength{\fboxsep}{1pt}%
    \definecolor{tbcol}{rgb}{1,1,1}%
\begin{picture}(7200.00,5040.00)%
    \gplgaddtomacro\gplbacktext{%
      \csname LTb\endcsname%
      \put(814,774){\makebox(0,0)[r]{\strut{}$0.4$}}%
      \csname LTb\endcsname%
      \put(814,1478){\makebox(0,0)[r]{\strut{}$0.6$}}%
      \csname LTb\endcsname%
      \put(814,2181){\makebox(0,0)[r]{\strut{}$0.8$}}%
      \csname LTb\endcsname%
      \put(814,2885){\makebox(0,0)[r]{\strut{}$1$}}%
      \csname LTb\endcsname%
      \put(814,3588){\makebox(0,0)[r]{\strut{}$1.2$}}%
      \csname LTb\endcsname%
      \put(814,4291){\makebox(0,0)[r]{\strut{}$1.4$}}%
      \csname LTb\endcsname%
      \put(1116,484){\makebox(0,0){\strut{}$0.6$}}%
      \csname LTb\endcsname%
      \put(1965,484){\makebox(0,0){\strut{}$0.7$}}%
      \csname LTb\endcsname%
      \put(2813,484){\makebox(0,0){\strut{}$0.8$}}%
      \csname LTb\endcsname%
      \put(3662,484){\makebox(0,0){\strut{}$0.9$}}%
      \csname LTb\endcsname%
      \put(4511,484){\makebox(0,0){\strut{}$1$}}%
      \csname LTb\endcsname%
      \put(5360,484){\makebox(0,0){\strut{}$1.1$}}%
      \csname LTb\endcsname%
      \put(6209,484){\makebox(0,0){\strut{}$1.2$}}%
    }%
    \gplgaddtomacro\gplfronttext{%
      \csname LTb\endcsname%
      \put(5816,4591){\makebox(0,0)[r]{\strut{}this work (3D)}}%
      \csname LTb\endcsname%
      \put(5816,4261){\makebox(0,0)[r]{\strut{}3D}}%
      \csname LTb\endcsname%
      \put(5816,3931){\makebox(0,0)[r]{\strut{}(2+1)D}}%
      \csname LTb\endcsname%
      \put(209,2761){\rotatebox{-270.00}{\makebox(0,0){\strut{}$\eta_\Phi$}}}%
      \put(3874,154){\makebox(0,0){\strut{}$\nu^{-1}$}}%
    }%
    \gplbacktext
    \put(0,0){\includegraphics[width={360.00bp},height={252.00bp}]{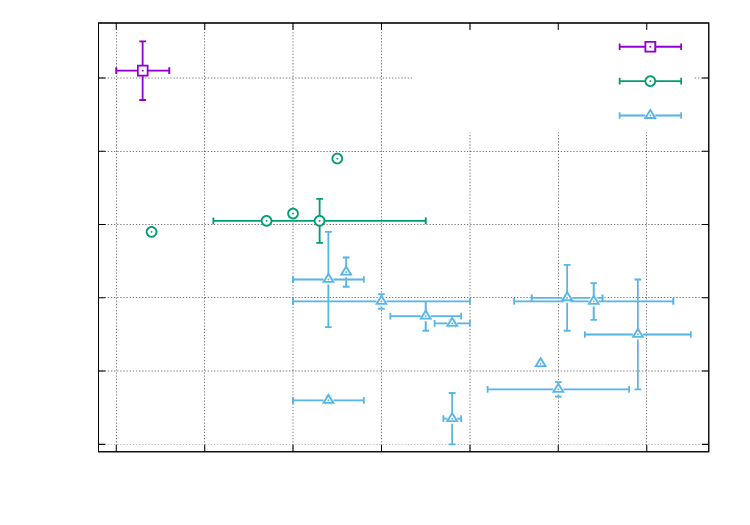}}%
    \gplfronttext
  \end{picture}%
\endgroup
 \caption{Critical exponents $\nu^{-1}$, $\eta_\Phi$ from Refs.~\cite{Ladovrechis:2022aof,Zerf:2017zqi,Gracey:2018qba,Knorr:2017yze,Janssen:2014gea,Lang:2025cwl,Lang:2018csk,Wang:2026jwf,Otsuka:2020lhc,Otsuka:2015iba,Liu:2021npk,Liu:2018sww,ParisenToldin:2014nkk,Xu:2020qbj,Ostmeyer:2021efs,Ostmeyer:2020uov,Buividovich:2018crq,Buividovich:2018yar} as listed in Tab.~\ref{tab:compendium}.
Points labelled ``\threeD'' (including this work) correspond to covariant methods with fully preserved Lorentz symmetry in 3 dimensions and are listed in the upper part of the table.
Points labelled ``\twoPoneD'' correspond to non-covariant methods that distinguish 2 spatial dimensions from one separate temporal dimension and are listed in the lower part of the table.
The absence of error bars indicates that no uncertainties were provided in the references, respectively.}
\label{fig:compendium} 
\end{figure}
Taken at face value our results suggest ${1\over\nu}=0.63(3)$,
$\eta_\Phi=2{\beta_\Phi\over\nu}-1=1.42(8)$, where 
we assume hyperscaling.
Tab.~\ref{tab:compendium} (based on Tab.~V of
\cite{Ladovrechis:2022aof}) presents a non-exhaustive list of estimates for these exponents obtained using
both
continuum-based approaches such as $(4-\varepsilon)$,
$(2+\varepsilon)$ and large-$N$ expansions and functional renormalisation group
(FRG), as well as lattice-based simulations such as  
determinantal Quantum Monte Carlo (QMC)~\cite{BSS:1981} and hybrid Monte Carlo (HMC)~\cite{Duane1987,Ostmeyer:2025agg}. The
discretisations admitting 8 relativistic fermionic components
in the low energy limit include both honeycomb and square lattices, the latter
with both $\pi$-flux lattice (equivalent to staggered fermions) and SLAC
discretisations. We have distinguished between approaches rooted in
\threeD\ covariant field theory (upper part), and those \twoPoneD\ simulations
in which the time direction is essentially
distinct (lower).

Significantly, our extracted results for both $\nu^{-1}$ and $\eta_\Phi$ 
are outliers, as becomes very apparent in Fig.~\ref{fig:compendium} where the values from Tab.~\ref{tab:compendium} are visualised.
Non-covariant \twoPoneD\ simulation-based
estimates for $\nu^{-1}$ tend to cluster around 1.0, and for $\eta_\Phi$ to be
significantly less than 1.0. Covariant \threeD\ approaches such as this one find smaller 
$\nu^{-1}$ and larger $\eta_\Phi$; even so our values lie at the
extremes of the range.

Fig.~\ref{fig:compendium} demonstrates that not only is there a very clear separation between the \threeD\ and \twoPoneD\ estimators, there is also a very clear overall anti-correlation between $\nu^{-1}$ and $\eta_\Phi$.
Our result is an outlier in both critical exponents, but it is reproducing this anti-correlation very well.
Generally, it appears to be relatively easy to estimate the sum $\nu^{-1}+\eta_\Phi$ (or something similar), but that the distinction between the two exponents is very hard.
Put differently, probably their (anti-)correlation is greatly underestimated in the uncertainties of most calculations (including ours).

\section{Fermion Correlator}\label{sec:fermions}
\subsection{Specification of Observable}
\label{sec:fermioncorrelator}
Here we introduce observables based on the fermion correlator 
\begin{equation}
{\cal S}(x,s;y,s^\prime)=\langle\Psi(x,s)\bar\Psi(y,s^\prime)\rangle.
\end{equation}
Following (\ref{eq:projection}), the \threeD\ propagator depends only on fields  
on the domain walls at $s=1,L_s$.
Define timeslice correlators~\cite{Hands:2022fhq}:
\begin{equation}
	\begin{split}
		S_\gamma(x_0)\equiv{1\over4}{\rm tr}\gamma_0S(x_0)&={1\over4}\sum_{\vec x}
		{\rm tr}\gamma_0[{\cal P}_-{\cal S}(0,1;x,1)+{\cal P}_+{\cal S}(0,L_s;x,L_s)]\,,\\
		S_m(x_0)\equiv{1\over4}{\rm tr}S(x_0)&={1\over4}\sum_{\vec x}
		{\rm tr}[{\cal P}_-{\cal S}(0,1;x,L_s)+{\cal P}_+{\cal S}(0,L_s;x,1)]\,,
	\end{split}
\label{eq:S}
\end{equation}
where the traces are taken over Dirac indices.
Empirically, the following relations are recovered on a generic
$\phi$ background in the
$L_s\to\infty$ limit:
\begin{eqnarray}
	\begin{split}
		(S_\gamma)_{11}=(S_\gamma)^*_{22}\,&;\;\;\;(S_\gamma)_{12}=-(S_\gamma)^*_{21}\,;\\
		(S_m)_{11}=-(S_m)^*_{22}\,&;\;\;\;(S_m)_{12}=(S_m)^*_{21},
	\end{split}
\label{eq:LargeLs}
\end{eqnarray}
where the subscripts are flavor indices, consistent with 
the global SU(2) symmetry (\ref{eq:SU(2)}).
For practical purposes it is thus sufficient to invert on just a single source flavor. 
For efficient sampling we implement 5 evenly spaced, but randomly shifted, wall sources~\cite{Hands:2022fhq}
for each $\phi$ configuration, performing 5
RHMC trajectories between measurements.
Our simulation then
reveal the only non-vanishing observables (up to noise) to be 
$\langle\Re(S_\gamma)_{11}\rangle$, $\langle\Re(S_m)_{11}\rangle$,
$\langle\Re(S_m)_{12}\rangle$, and $\langle\Im(S_m)_{12}\rangle$
as well as those related by equation~\eqref{eq:LargeLs};
below we will label the three $S_m$ components by an iso-index $a=3,1,2$ respectively.

We find the flavor-singlet component $S_\gamma$ is
insensitive to the coupling $\beta$. Just as for the GN
model~\cite{Hands:2016foa}, $S_\gamma$ is symmetric under Euclidean
time reversal.  
By contrast $S_m$ is odd under Euclidean time reversal, again echoing the 
GN model~\cite{Hands:2016foa}. This time we find a clear
$\beta$-dependence, so we focus subsequent analysis on the $S_m$ component.

Since there is no symmetry-breaking term in the simulation, $\vec\phi$ 
drifts uniformly around its SO(3) manifold during the course of the
simulation, as shown in Fig.~\ref{fig:drift}, 
so none of the three non-vanishing components of $S_m$ should be
privileged over sufficently long run times. Noting that in the $L_s\to\infty$
limit (\ref{eq:LargeLs}) implies $S_m(x_0)=\vec\tau\cdot \vec Q$, our analysis
first applies a global rotation $S_m\to\tilde S_m$ such that
$\vec Q(x_0=1)=(0,0,Q_3)$, in effect rotating the source along the third
iso-direction. Further details are given in Appendix~\ref{app:B}.

\subsection{THC fit results}
\label{sec:THC}

\begin{figure}[t] 
	\centering
	\includegraphics[width=.32\textwidth,page=261,trim=0 1cm 1cm 2cm, clip]{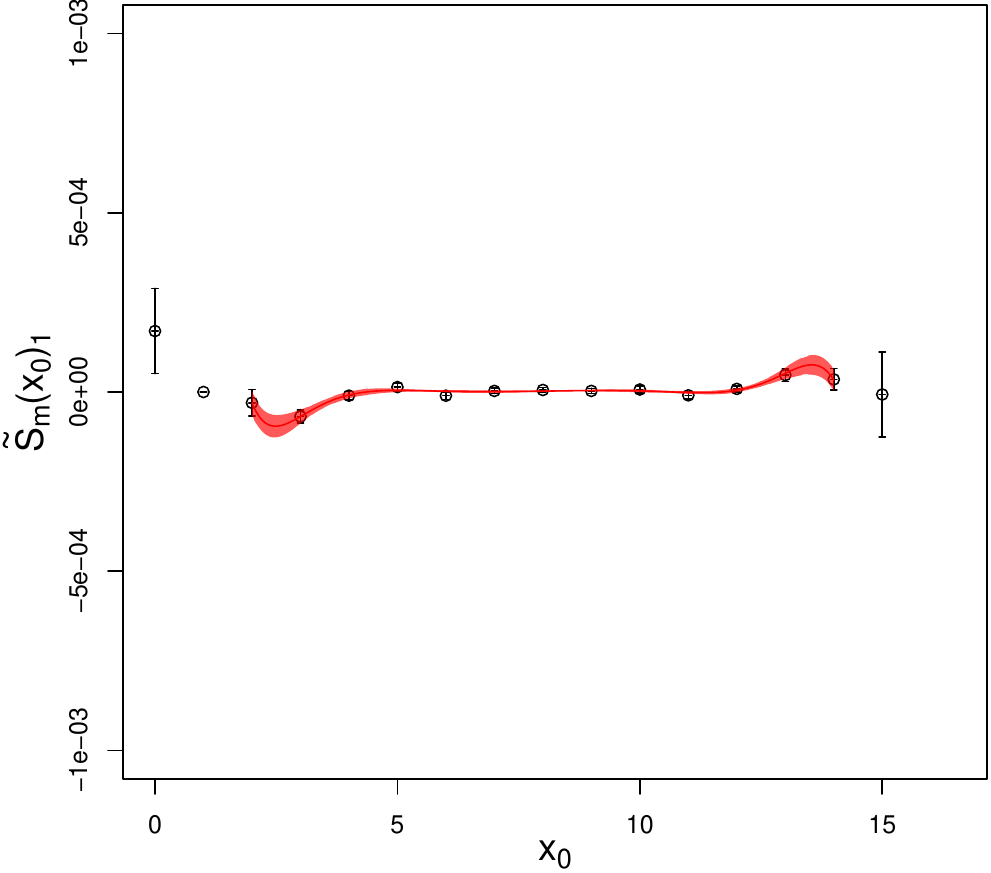} 
	\hfill
	\includegraphics[width=.32\textwidth,page=262,trim=0 1cm 1cm 2cm, clip]{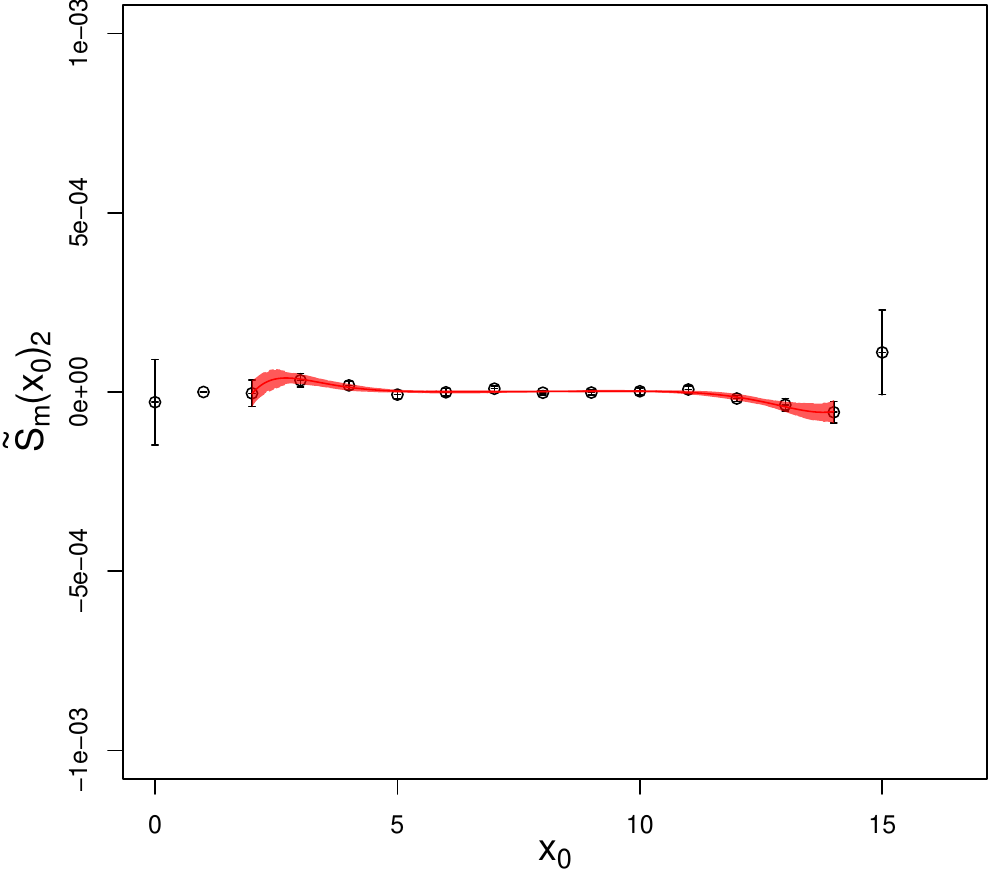} 
	\hfill
	\includegraphics[width=.32\textwidth,page=260,trim=0 1cm 1cm 2cm, clip]{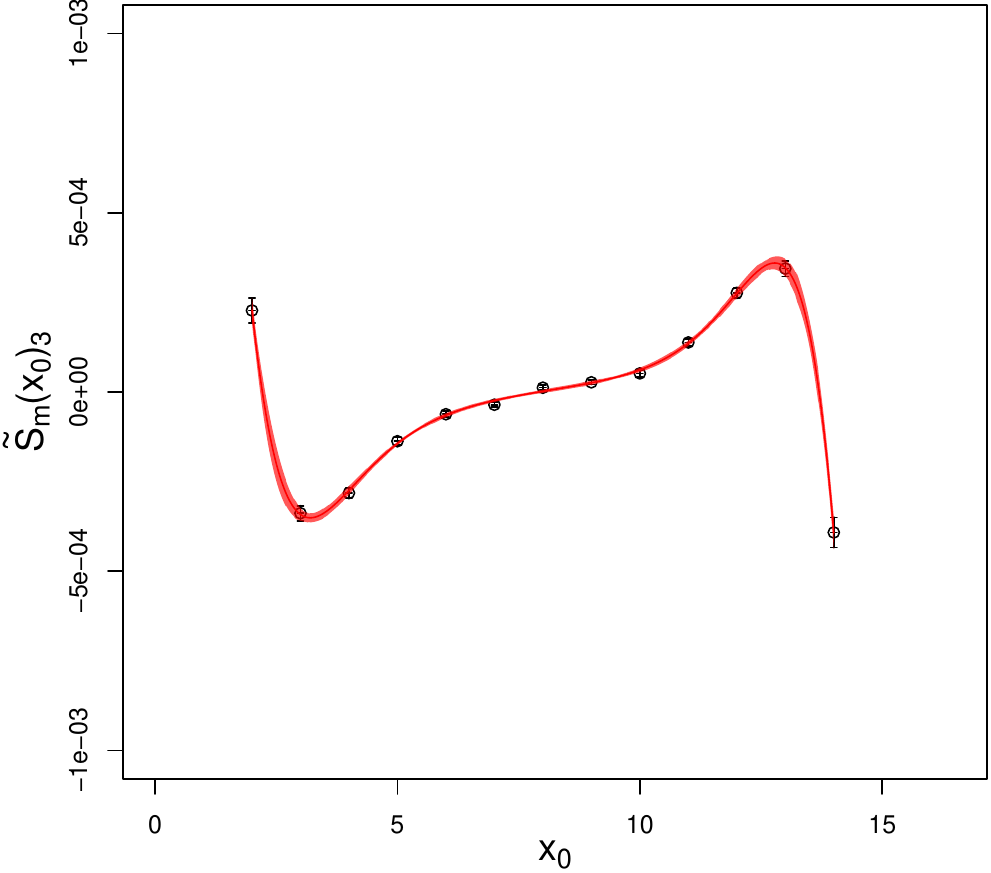} 
	\caption{Expectation values $\tilde S_m(x_0)$ of the $a=1,2,3$-components (left
		to right) of the fermion correlator after rotation of the time slice $t=1$ into
		positive $3$-direction for $L=16$, $L_s=24$, coupling $\beta=\num{0.375}$.  The
		red curves show a simultaneous 8-state fit to the $2\times 2$ matrix $\tilde S_m$
		including error bands obtained with the THC method.} \label{fig:thc-fit-2x2}
\end{figure}

After rotation we fit the full $2\times 2$ matrix-valued fermion correlator
$\tilde S_m(x_0)$.  This simultaneous fit is made feasible in an automated way by the
Truncated Hankel Correlator (THC) method~\cite{Ostmeyer:2025igc}.  The THC is a
fully algebraic method providing a near-optimal set of energies $E_l$ and matrix
elements $c_l$ for any given truncation $k$ approximating \begin{align}
\br{\tilde S_m(x_0)}_{\alpha\beta} &= \sum_{l=1}^{k} c_{\alpha\beta,l}\, \eto{-E_l
x_0} \end{align} with a tower of exponentials.  Anti-symmetry in this setting is
realised by pairs of energies $E_l=-E_{l'}$ and coefficients
$c_{\alpha\beta,l}=- \eto{E_{l} L} c_{\alpha\beta,l'}$.
\mbox{(Anti-)symmetries} of
this type can be captured exactly by the THC method without the necessity to
constrain the parameters {\em a priori\/}.  We use the implementation in the
\texttt{hadron} package~\cite{hadron}.

\begin{figure}[t] 
	\centering
	\includegraphics[width=.48\textwidth,page=192,trim=0 1cm 1cm 2cm, clip]{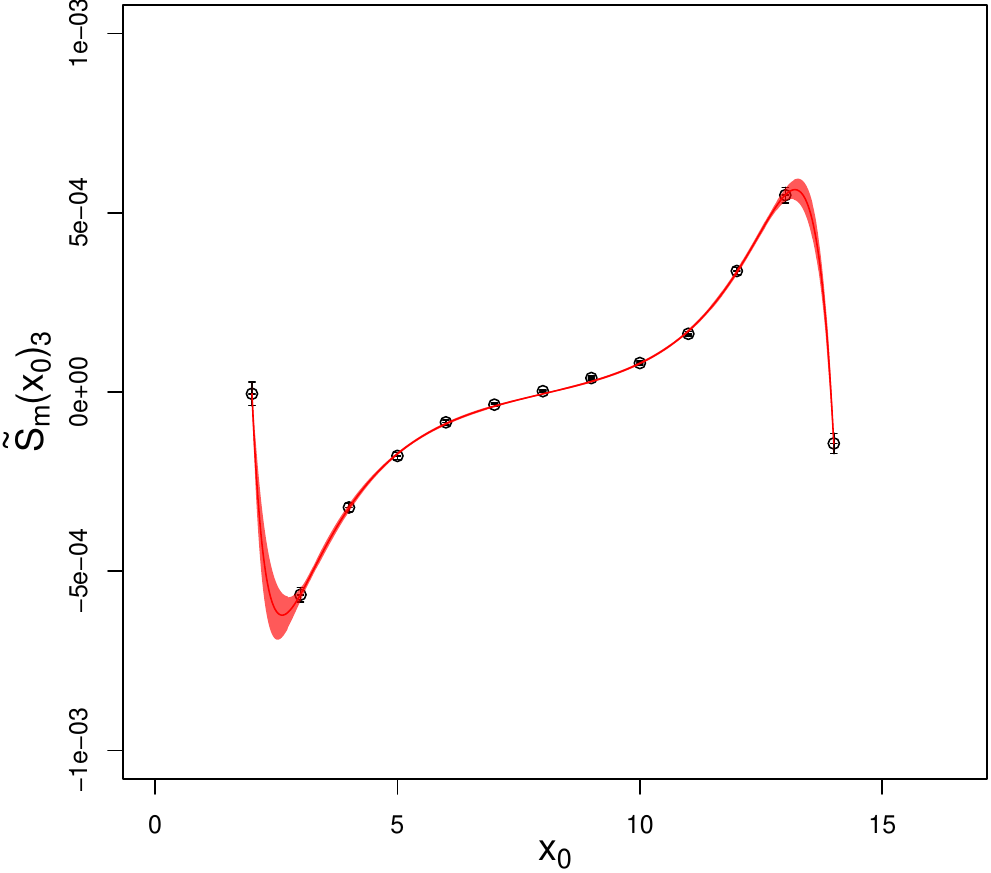}
	\hfill
	\includegraphics[width=.48\textwidth,page=237,trim=0 1cm 1cm 2cm, clip]{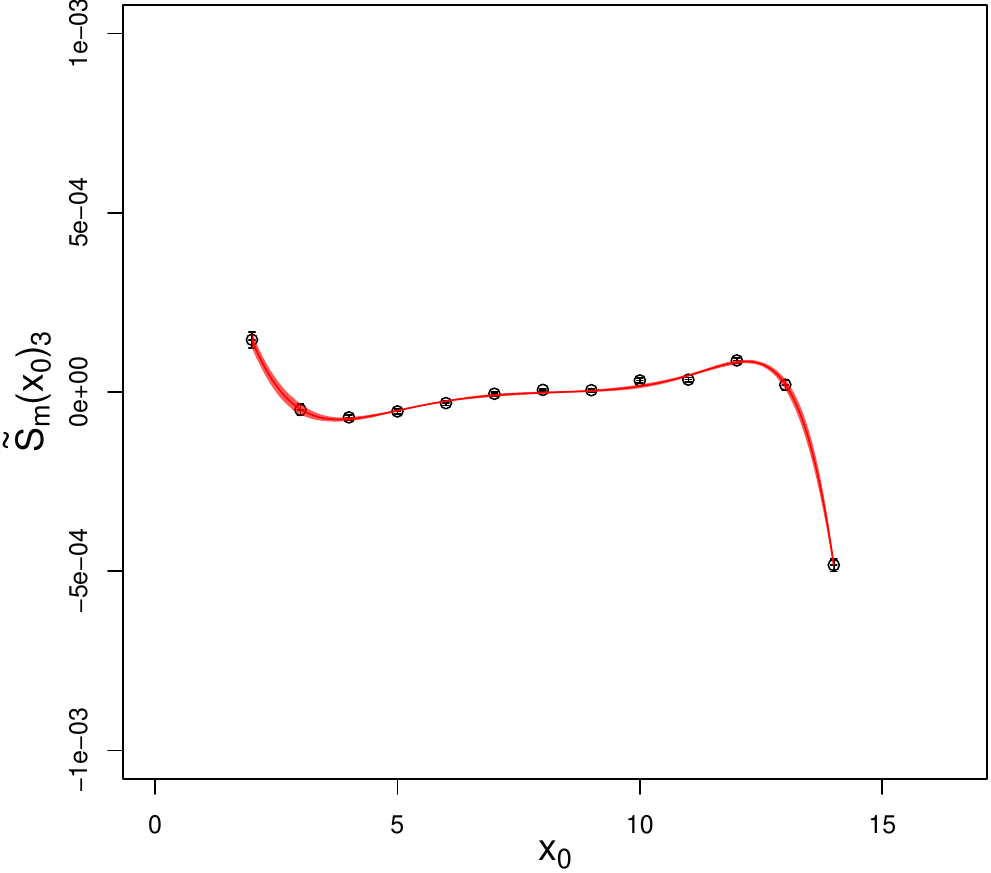} 
	\caption{Expectation value $\tilde S_m(x_0)$ of the $3$-component of the fermion correlator after rotation of the time slice $x_0=1$ into positive $3$-direction.
		Both panels used $L=L_s=16$.
		Left: coupling $\beta=\num{0.35}$, right: $\beta=\num{0.475}$.
		The red curves show 4-state fits including error bands obtained with the THC method.}
	\label{fig:thc-fit} 
\end{figure}

An example of such a matrix-valued THC fit is visualised in
Fig.~\ref{fig:thc-fit-2x2}.  In this case, we required $k=8$ states to
adequately describe the behaviour for $2\le x_0\le L-2$.  The correlator is
anti-symmetric to a very good approximation which means that the data is
effectively described by 4 hyperbolic sine functions \begin{align}
\br{\tilde S_m(x_0)}_{\alpha\beta} &= \sum_{l=1}^{k/2} \tilde c_{\alpha\beta,l}\,
\sinh\br{E_l (x_0-L/2)}\,.  \end{align} 
The most important result of
all our simultaneous fits is that the $\{\alpha,\beta\}=\{1,2\}$-components are always compatible
with zero.  In other words, once the gauge-fixing rotation aligns the
correlators at $x_0=1$ into the $3$-direction, the entire fermion correlator remains
within this 1-dimensional subspace up to statistical fluctuations.
Physically, this confirms  that the residual U$(1)$ acting in the $1$$2$-plane following SU(2) symmetry
breaking 
remains unbroken as expected.

\begin{figure}[t] 
	\centering
\begingroup
  \inputencoding{latin1}%
  \makeatletter
  \providecommand\color[2][]{%
    \GenericError{(gnuplot) \space\space\space\@spaces}{%
      Package color not loaded in conjunction with
      terminal option `colourtext'%
    }{See the gnuplot documentation for explanation.%
    }{Either use 'blacktext' in gnuplot or load the package
      color.sty in LaTeX.}%
    \renewcommand\color[2][]{}%
  }%
  \providecommand\includegraphics[2][]{%
    \GenericError{(gnuplot) \space\space\space\@spaces}{%
      Package graphicx or graphics not loaded%
    }{See the gnuplot documentation for explanation.%
    }{The gnuplot epslatex terminal needs graphicx.sty or graphics.sty.}%
    \renewcommand\includegraphics[2][]{}%
  }%
  \providecommand\rotatebox[2]{#2}%
  \@ifundefined{ifGPcolor}{%
    \newif\ifGPcolor
    \GPcolortrue
  }{}%
  \@ifundefined{ifGPblacktext}{%
    \newif\ifGPblacktext
    \GPblacktexttrue
  }{}%
  \let\gplgaddtomacro\g@addto@macro
  \gdef\gplbacktext{}%
  \gdef\gplfronttext{}%
  \makeatother
  \ifGPblacktext
    \def\colorrgb#1{}%
    \def\colorgray#1{}%
  \else
    \ifGPcolor
      \def\colorrgb#1{\color[rgb]{#1}}%
      \def\colorgray#1{\color[gray]{#1}}%
      \expandafter\def\csname LTw\endcsname{\color{white}}%
      \expandafter\def\csname LTb\endcsname{\color{black}}%
      \expandafter\def\csname LTa\endcsname{\color{black}}%
      \expandafter\def\csname LT0\endcsname{\color[rgb]{1,0,0}}%
      \expandafter\def\csname LT1\endcsname{\color[rgb]{0,1,0}}%
      \expandafter\def\csname LT2\endcsname{\color[rgb]{0,0,1}}%
      \expandafter\def\csname LT3\endcsname{\color[rgb]{1,0,1}}%
      \expandafter\def\csname LT4\endcsname{\color[rgb]{0,1,1}}%
      \expandafter\def\csname LT5\endcsname{\color[rgb]{1,1,0}}%
      \expandafter\def\csname LT6\endcsname{\color[rgb]{0,0,0}}%
      \expandafter\def\csname LT7\endcsname{\color[rgb]{1,0.3,0}}%
      \expandafter\def\csname LT8\endcsname{\color[rgb]{0.5,0.5,0.5}}%
    \else
      \def\colorrgb#1{\color{black}}%
      \def\colorgray#1{\color[gray]{#1}}%
      \expandafter\def\csname LTw\endcsname{\color{white}}%
      \expandafter\def\csname LTb\endcsname{\color{black}}%
      \expandafter\def\csname LTa\endcsname{\color{black}}%
      \expandafter\def\csname LT0\endcsname{\color{black}}%
      \expandafter\def\csname LT1\endcsname{\color{black}}%
      \expandafter\def\csname LT2\endcsname{\color{black}}%
      \expandafter\def\csname LT3\endcsname{\color{black}}%
      \expandafter\def\csname LT4\endcsname{\color{black}}%
      \expandafter\def\csname LT5\endcsname{\color{black}}%
      \expandafter\def\csname LT6\endcsname{\color{black}}%
      \expandafter\def\csname LT7\endcsname{\color{black}}%
      \expandafter\def\csname LT8\endcsname{\color{black}}%
    \fi
  \fi
    \setlength{\unitlength}{0.0500bp}%
    \ifx\gptboxheight\undefined%
      \newlength{\gptboxheight}%
      \newlength{\gptboxwidth}%
      \newsavebox{\gptboxtext}%
    \fi%
    \setlength{\fboxrule}{0.5pt}%
    \setlength{\fboxsep}{1pt}%
    \definecolor{tbcol}{rgb}{1,1,1}%
\begin{picture}(7200.00,5040.00)%
    \gplgaddtomacro\gplbacktext{%
      \csname LTb\endcsname%
      \put(814,704){\makebox(0,0)[r]{\strut{}$0$}}%
      \csname LTb\endcsname%
      \put(814,1390){\makebox(0,0)[r]{\strut{}$0.2$}}%
      \csname LTb\endcsname%
      \put(814,2076){\makebox(0,0)[r]{\strut{}$0.4$}}%
      \csname LTb\endcsname%
      \put(814,2762){\makebox(0,0)[r]{\strut{}$0.6$}}%
      \csname LTb\endcsname%
      \put(814,3447){\makebox(0,0)[r]{\strut{}$0.8$}}%
      \csname LTb\endcsname%
      \put(814,4133){\makebox(0,0)[r]{\strut{}$1$}}%
      \csname LTb\endcsname%
      \put(814,4819){\makebox(0,0)[r]{\strut{}$1.2$}}%
      \csname LTb\endcsname%
      \put(946,484){\makebox(0,0){\strut{}$0.34$}}%
      \csname LTb\endcsname%
      \put(1635,484){\makebox(0,0){\strut{}$0.36$}}%
      \csname LTb\endcsname%
      \put(2324,484){\makebox(0,0){\strut{}$0.38$}}%
      \csname LTb\endcsname%
      \put(3013,484){\makebox(0,0){\strut{}$0.4$}}%
      \csname LTb\endcsname%
      \put(3702,484){\makebox(0,0){\strut{}$0.42$}}%
      \csname LTb\endcsname%
      \put(4391,484){\makebox(0,0){\strut{}$0.44$}}%
      \csname LTb\endcsname%
      \put(5080,484){\makebox(0,0){\strut{}$0.46$}}%
      \csname LTb\endcsname%
      \put(5769,484){\makebox(0,0){\strut{}$0.48$}}%
      \csname LTb\endcsname%
      \put(6458,484){\makebox(0,0){\strut{}$0.5$}}%
    }%
    \gplgaddtomacro\gplfronttext{%
      \csname LTb\endcsname%
      \put(1870,1592){\makebox(0,0)[r]{\strut{}  $L=16$}}%
      \csname LTb\endcsname%
      \put(1870,1262){\makebox(0,0)[r]{\strut{}  $L=20$}}%
      \csname LTb\endcsname%
      \put(1870,932){\makebox(0,0)[r]{\strut{}  $L=24$}}%
      \csname LTb\endcsname%
      \put(209,2761){\rotatebox{-270.00}{\makebox(0,0){\strut{}$E_0$}}}%
      \put(3874,154){\makebox(0,0){\strut{}$\beta$}}%
    }%
    \gplbacktext
    \put(0,0){\includegraphics[width={360.00bp},height={252.00bp}]{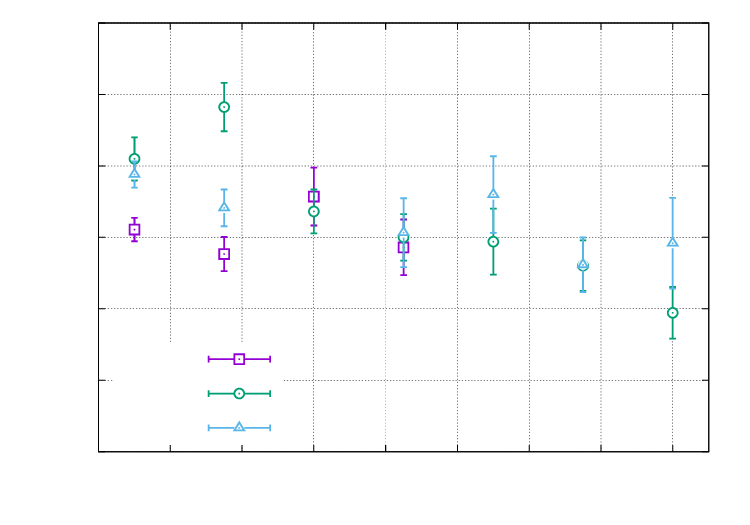}}%
    \gplfronttext
  \end{picture}%
\endgroup
	\caption{Ground state energies $E_0$ of the fermion correlators as a
		function of the coupling $\beta$ for $L_s=16$.
		The plots look very similar for $L_s=8,24$.}
	\label{fig:energies} 
\end{figure}

Once the $3$-component has been isolated and numerically verified to be the only
significant signal, we can simplify the analysis and fit this component
exclusively.  Fit results for two very different values of the coupling $\beta$
are depicted in Fig.~\ref{fig:thc-fit}.  Both cases are described well by 4
exponential or 2 sinh states.  Their major difference is the greatly decreased
amplitude for larger~$\beta$.

Surprisingly, the quantification of this difference is rather elusive.  The
ground state energy $E_0$ remains approximately constant over the entire region
of $\beta$ as can be seen in Fig.~\ref{fig:energies}.  That is, the fermion
remains massive across the phase transition and only its interaction
with the field changes.

\begin{figure}[t] 
	\centering
\begingroup
  \inputencoding{latin1}%
  \makeatletter
  \providecommand\color[2][]{%
    \GenericError{(gnuplot) \space\space\space\@spaces}{%
      Package color not loaded in conjunction with
      terminal option `colourtext'%
    }{See the gnuplot documentation for explanation.%
    }{Either use 'blacktext' in gnuplot or load the package
      color.sty in LaTeX.}%
    \renewcommand\color[2][]{}%
  }%
  \providecommand\includegraphics[2][]{%
    \GenericError{(gnuplot) \space\space\space\@spaces}{%
      Package graphicx or graphics not loaded%
    }{See the gnuplot documentation for explanation.%
    }{The gnuplot epslatex terminal needs graphicx.sty or graphics.sty.}%
    \renewcommand\includegraphics[2][]{}%
  }%
  \providecommand\rotatebox[2]{#2}%
  \@ifundefined{ifGPcolor}{%
    \newif\ifGPcolor
    \GPcolortrue
  }{}%
  \@ifundefined{ifGPblacktext}{%
    \newif\ifGPblacktext
    \GPblacktexttrue
  }{}%
  \let\gplgaddtomacro\g@addto@macro
  \gdef\gplbacktext{}%
  \gdef\gplfronttext{}%
  \makeatother
  \ifGPblacktext
    \def\colorrgb#1{}%
    \def\colorgray#1{}%
  \else
    \ifGPcolor
      \def\colorrgb#1{\color[rgb]{#1}}%
      \def\colorgray#1{\color[gray]{#1}}%
      \expandafter\def\csname LTw\endcsname{\color{white}}%
      \expandafter\def\csname LTb\endcsname{\color{black}}%
      \expandafter\def\csname LTa\endcsname{\color{black}}%
      \expandafter\def\csname LT0\endcsname{\color[rgb]{1,0,0}}%
      \expandafter\def\csname LT1\endcsname{\color[rgb]{0,1,0}}%
      \expandafter\def\csname LT2\endcsname{\color[rgb]{0,0,1}}%
      \expandafter\def\csname LT3\endcsname{\color[rgb]{1,0,1}}%
      \expandafter\def\csname LT4\endcsname{\color[rgb]{0,1,1}}%
      \expandafter\def\csname LT5\endcsname{\color[rgb]{1,1,0}}%
      \expandafter\def\csname LT6\endcsname{\color[rgb]{0,0,0}}%
      \expandafter\def\csname LT7\endcsname{\color[rgb]{1,0.3,0}}%
      \expandafter\def\csname LT8\endcsname{\color[rgb]{0.5,0.5,0.5}}%
    \else
      \def\colorrgb#1{\color{black}}%
      \def\colorgray#1{\color[gray]{#1}}%
      \expandafter\def\csname LTw\endcsname{\color{white}}%
      \expandafter\def\csname LTb\endcsname{\color{black}}%
      \expandafter\def\csname LTa\endcsname{\color{black}}%
      \expandafter\def\csname LT0\endcsname{\color{black}}%
      \expandafter\def\csname LT1\endcsname{\color{black}}%
      \expandafter\def\csname LT2\endcsname{\color{black}}%
      \expandafter\def\csname LT3\endcsname{\color{black}}%
      \expandafter\def\csname LT4\endcsname{\color{black}}%
      \expandafter\def\csname LT5\endcsname{\color{black}}%
      \expandafter\def\csname LT6\endcsname{\color{black}}%
      \expandafter\def\csname LT7\endcsname{\color{black}}%
      \expandafter\def\csname LT8\endcsname{\color{black}}%
    \fi
  \fi
    \setlength{\unitlength}{0.0500bp}%
    \ifx\gptboxheight\undefined%
      \newlength{\gptboxheight}%
      \newlength{\gptboxwidth}%
      \newsavebox{\gptboxtext}%
    \fi%
    \setlength{\fboxrule}{0.5pt}%
    \setlength{\fboxsep}{1pt}%
    \definecolor{tbcol}{rgb}{1,1,1}%
\begin{picture}(7200.00,5040.00)%
    \gplgaddtomacro\gplbacktext{%
      \csname LTb\endcsname%
      \put(814,704){\makebox(0,0)[r]{\strut{}$0$}}%
      \csname LTb\endcsname%
      \put(814,1390){\makebox(0,0)[r]{\strut{}$100$}}%
      \csname LTb\endcsname%
      \put(814,2076){\makebox(0,0)[r]{\strut{}$200$}}%
      \csname LTb\endcsname%
      \put(814,2762){\makebox(0,0)[r]{\strut{}$300$}}%
      \csname LTb\endcsname%
      \put(814,3447){\makebox(0,0)[r]{\strut{}$400$}}%
      \csname LTb\endcsname%
      \put(814,4133){\makebox(0,0)[r]{\strut{}$500$}}%
      \csname LTb\endcsname%
      \put(814,4819){\makebox(0,0)[r]{\strut{}$600$}}%
      \csname LTb\endcsname%
      \put(946,484){\makebox(0,0){\strut{}$0.34$}}%
      \csname LTb\endcsname%
      \put(1635,484){\makebox(0,0){\strut{}$0.36$}}%
      \csname LTb\endcsname%
      \put(2324,484){\makebox(0,0){\strut{}$0.38$}}%
      \csname LTb\endcsname%
      \put(3013,484){\makebox(0,0){\strut{}$0.4$}}%
      \csname LTb\endcsname%
      \put(3702,484){\makebox(0,0){\strut{}$0.42$}}%
      \csname LTb\endcsname%
      \put(4391,484){\makebox(0,0){\strut{}$0.44$}}%
      \csname LTb\endcsname%
      \put(5080,484){\makebox(0,0){\strut{}$0.46$}}%
      \csname LTb\endcsname%
      \put(5769,484){\makebox(0,0){\strut{}$0.48$}}%
      \csname LTb\endcsname%
      \put(6458,484){\makebox(0,0){\strut{}$0.5$}}%
    }%
    \gplgaddtomacro\gplfronttext{%
      \csname LTb\endcsname%
      \put(5816,4591){\makebox(0,0)[r]{\strut{}  $L=12$}}%
      \csname LTb\endcsname%
      \put(5816,4261){\makebox(0,0)[r]{\strut{}  $L=16$}}%
      \csname LTb\endcsname%
      \put(5816,3931){\makebox(0,0)[r]{\strut{}  $L=20$}}%
      \csname LTb\endcsname%
      \put(5816,3601){\makebox(0,0)[r]{\strut{}  $L=24$}}%
      \csname LTb\endcsname%
      \put(209,2761){\rotatebox{-270.00}{\makebox(0,0){\strut{}$L^6 \tilde c_0 E_0$}}}%
      \put(3874,154){\makebox(0,0){\strut{}$\beta$}}%
    }%
    \gplbacktext
    \put(0,0){\includegraphics[width={360.00bp},height={252.00bp}]{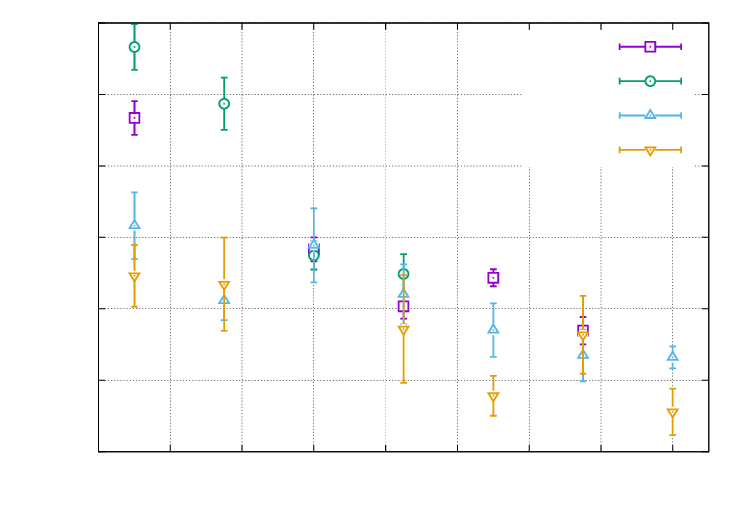}}%
    \gplfronttext
  \end{picture}%
\endgroup
	\caption{Ground state contribution to the slope $\tilde c_0 E_0$ of the correlator at the zero crossing at $x_0=L/2$ rescaled by the squared volume $V=L^3$ as a function of the coupling $\beta$ for $L_s=16$.
		The plots look very similar for $L_s=8,24$.}
	\label{fig:slopes} 
\end{figure}

An alternative observable related to the amplitude of the correlator is the
ground state contribution to the slope of the correlator at half time $\tilde
c_0 E_0$.  We see the amplitude decrease in Fig.~\ref{fig:thc-fit} reflected
in a decrease of this slope in Fig.~\ref{fig:slopes}.
Since the slope decreases equally quickly for
all lattice sizes, it does not feature a phase transition signature either.
The leading factor
appears to be inversely proportional to the squared volume $V=L^3$ and has been
included in Fig.~\ref{fig:slopes} to achieve an approximate data collapse.
This need have little to do with the underlying fermion
propagation and is conceivably due to the decreased sampling efficiency of the global
rotation $S_m\to\tilde S_m$ as the spatial extent of correlated patches of
$\phi$-field decreases with increasing $\beta$.
However, given the considerable finite volume artifacts for free DWF discussed in~\cite{Hands:2022fhq},
we should be wary of drawing firm conclusions from relatively small systems.
 
\section{Summary}

\label{sec:summary}

We have investigated the SU$(2)$ chiral Heisenberg model as a \threeD\ covariant
lattice field theory of Gross-Neveu (GN) type.  It lends its name to a
well-populated universality class of phase transitions.  Yet, to date lattice
simulations of other models within the chiral Heisenberg universality class were
limited to \twoPoneD, that is they have a distinct (non-covariant) time
dimension.  This includes, but is not limited to, the Hubbard model on honeycomb
and $\pi$-flux lattices.  GN models, on the other hand, have primarily been
investigated perturbatively so far.  In contrast, in this work we have performed
a non-perturbative DWF simulation using the 
RHMC algorithm.

We have treated the absolute value $|\Phi|$ of the volume-averaged 3-vector
$\vec \phi$ that encodes the bosonic fields as an order parameter.  At a
critical coupling $\beta_c$ the SU$(2)$ symmetry is broken spontaneously to a
residual U$(1)$ and $|\Phi|$ develops a non-zero expectation value for all
smaller couplings $\beta<\beta_c$.  First, we have determined the Binder
cumulant $B$ (see Fig.~\ref{fig:binder}) and observed simultaneous a crossing of
$B$ for different volumes.  While this is a clear signature of a phase
transition, we found the Binder cumulant too noisy for a quantitative analysis.
Instead, we performed a data collapse fit assuming universal finite volume
scaling (FVS), see Fig.-\ref{fig:plot5_Ls24}.  The FVS
ansatz allowed us to extract the critical coupling $\beta_c=0.465(11)$ as well
as the critical exponents ${\beta_\Phi\over\nu}=1.21(4)$ and
${1\over\nu}=0.633(26)$.
The numerical fits underpinning these results are stable and in particularly show no sensitivity
to the domain wall separation $L_s$, suggesting that the $L_s\to\infty$ limit
needed to recover the symmetries of fully relativitic fermions is readily
reached in this model, corroborating earlier findings in the \threeD~GN model
~\cite{Hands:2016foa}.

Our result for $\beta_c$ is uncontroversial, but our
critical exponents are clear outliers compared to practically all findings in
earlier work.  We have visualised 18 distinct tuples
$({1\over\nu},{\beta_\Phi\over\nu})$ from various literature sources in
Fig.~\ref{fig:compendium} together with our new values. It would also be very
interesting to compare with anticipated results from the conformal bootstrap.
The majority of
literature values are not mutually compatible, however, there are two trends
apparent.  First of all, the results are completely split into two almost
disjoint clusters, one stemming from methods using \threeD\ models (including
ours) and another based on \twoPoneD\ models.  The second trend is a strong
anti-correlation between ${1\over\nu}$ and ${\beta_\Phi\over\nu}$ over all the
results.  Our understanding of this anti-correlation is that the two critical
exponents are very hard to distinguish methodologically and that systematic
uncertainties that result from this difficulty are vastly underestimated in
(almost) all determinations including our own.  We encourage the community to
contemplate this behaviour in the hope to pin down the characteristics of the
chiral Heisenberg universality class conclusively.

Subsequently, in Sec.~\ref{sec:fermions} we have investigated the fermion
correlator.  Since we did not introduce any explicit symmetry breaking or bare
fermion mass to our simulations, the correlator is a hermitian $2\times 2$
matrix with the full SU$(2)$ symmetry preserved and anti-symmetric in time.
This made the analysis non-amenable to standard techniques like plateau fits.
To obtain a non-vanishing signal, we needed to rotate each ensemble into a fixed
reference frame prior to averaging.  A fit to the full matrix-valued correlator
with the Truncated Hankel Correlator (THC) method~\cite{Ostmeyer:2025igc} then
allowed to describe the correlator with a tower of sinh functions.  It showed
the expected unbroken U$(1)$ sub-symmetry on both sides of the phase transition.
We further found that the ground state energy hardly depends on the coupling
$\beta$ which demonstrates that fermion remains massive across the phase
transition.  Its interaction with the bosonic fields, on the other hand, changes
with $\beta$, reflected in a decreasing amplitude of the correlator (see
Figs.~\ref{fig:thc-fit},\ref{fig:slopes}).  The fermionic correlator data do
not feature a phase transition signature which certainly warrants further
investigation in future work on larger lattices.
 
\section*{Acknowledgements}
We've enjoyed helpful discussions with John Gracey, Igor Herbut and Michael Scherer.
This work was performed using the Cambridge Service for Data Driven Discovery
(CSD3), part of which is operated by the University of Cambridge Research
Computing on behalf of the STFC DiRAC HPC Facility (www.dirac.ac.uk). The DiRAC
component of CSD3 was funded by BEIS capital funding via STFC capital grants
ST/P002307/1 and ST/R002452/1 and STFC operations grant ST/R00689X/1. DiRAC is
part of the National e-Infrastructure. 
SJH was supported in part by the STFC Consolidated Grant ST/T000813/1, and JO by ST/T000988/1
as well as the Deutsche Forschungsgemeinschaft (DFG, German Research Foundation) as part of the
CRC 1639 NuMeriQS – Project number 511713970.

\section*{Data Access} 
The data (raw as well as analysed) required
to reproduce the results of this paper is published under open
access~\cite{liverpool_rdm3076}. We used \texttt{Fortran} for the DWF simulations~\cite{hands_2026_19262097}.
Our data analysis relied heavily on the standalone package \texttt{comp-avg}~\cite{comp-avg} as well as
the \texttt{hadron}~\cite{hadron} package implemented in \texttt{R}~\cite{r_language}.

\appendix
\section{Sign of the Determinant}
\label{app:A}
In the continuum Lagrangian density (\ref{eq:Lcont}) the fermion matrix ${\cal
M}$ is the sum of an anti-hermitian and a
hermitian term, so its eigenvalues are complex in general. 
If,
following a unitary rotation on $\psi$~\cite{Hands:2015qha},
we replace the interaction term by the U(2)-equivalent form
$g\vec\phi\cdot \bar\psi i\gamma_3\otimes\vec\tau\psi$ then both terms are anti-hermitian, and the
spectrum of ${\cal M}$ is pure imaginary. Since ${\cal M}$ is even-dimensional, this
proves the determinant $\mbox{det}{\cal M}$ is real. 
The more conventional
route uses the identity ${\cal M}^\dagger=\gamma_5 {\cal M}\gamma_5$ to show
$\mbox{det}{\cal M}=(\mbox{det}{\cal M})^*$ This is important because this proof extends to
lattice regularisations such as DWF. We will
proceed on the assumption that $\mbox{det}{\cal M}$ is real, for both target and
regularised theories.

Theories with a real determinant can still have a sign problem, but its nature is
significantly constrained. Eigenvalues of ${\cal M}$ either come as complex $(\lambda,
\lambda^*)$ pairs or lie on the real axis. If the real eigenvalues can move
independently along the axis, then it is possible for an odd number to become
negative, flipping the sign of $\mbox{det}{\cal M}$. 
Theories with isolated real eigenvalues are
interesting because under smooth evolution (eg.\ under HMC flow) the eigenvalue
must remain on the real axis until it meets another isolated eigenvalue,
whereupon both have the opportunity to move back into the plane as part of a 
conjugate pair.

In order to see which case applies here, 
write the eigenvalue equation:
\begin{equation}
[\partial_\mu\gamma_\mu\otimes\One+i\vec\phi\cdot \gamma_3\otimes\vec\tau]\Psi=\lambda\Psi\,.
\end{equation}
Now consider a related mode $\tilde\Psi={\cal C}\gamma_5\otimes\tau_2\Psi$, where the (Dirac-basis
dependent) matrix
${\cal C}$ has the properties
\begin{equation}
{\cal C}\gamma_\mu{\cal C}^{-1}=-\gamma_\mu^*;\;\;\;{\cal C}^{-1}=-{\cal
C};\;\;\;{\cal C}^*={\cal C};\;\;\;[{\cal C},\gamma_5]=0,
\end{equation}
satisfying
\begin{equation}
[\partial_\mu\gamma_\mu^*\otimes\One-i\vec\phi\cdot \gamma_3^*\otimes\vec\tau^*]\tilde\Psi=\lambda\tilde\Psi,
\end{equation}
where we have also used $\tau_2\vec\tau\tau_2=-\vec\tau^*$. We conclude
$\tilde\Psi^*$ is another eigenmode of ${\cal M}$ with eigenvalue $\lambda^*=-\lambda$.
Since every non-zero eigenvalue is a member of a conjugate pair, 
$\mbox{det}{\cal M}$ is real and positive. 
We can develop an analogous argument using
the hermitian interaction term if instead we focus on the hermitian matrix
$\gamma_5{\cal M}$, to show that all eigenvalues are real and that both $\lambda$ and
$-\lambda$ feature in the spectrum, with corresponding modes ($\Psi,\,{\cal
C}\otimes\tau_2\Psi^*$). Since $\mbox{det}\gamma_5=1$, and by construction
${\rm dim}({\cal M})$ is a multiple of 4, this is enough
to prove the same result.

Things get interesting when we go to the lattice regularisation since  
DWF contain a hermitian term in ${\cal M}$ proportional to
$\One\otimes\One$; therefore   $[\gamma_5\otimes\One,{\cal
C}\otimes\tau_2]=0$ spoils the $\lambda\to-\lambda$ symmetry for
$\gamma_5{\cal M}$. The mode
$\tilde\Psi^*$ satisfies a different equation to
$\Psi$, we cannot deduce spectral symmetry, and we conclude $\mbox{det}{\cal M}$ is not positive in general.
Since we expect to recover the target theory in the limit $L_s\to\infty$, it
seems reasonable to hope that the resulting sign problem is mild and can
be ignored.

\section{Gauge fixing (rotations into $z$-basis)}
\label{app:B}
Let us consider an observable $S$, for instance $S_m$ from equation~\eqref{eq:LargeLs}, of the form
\begin{align}
	S &= \matr{cc}{x_3 & x_1 - \im x_2\\x_1 + \im x_2 & -x_3}\\
	&\equiv x_k \sigma_k\,, \label{eq:S_as_3-vector}
\end{align}
i.e.\ $S$ is a traceless hermitian matrix that can be represented by a 3-vector
$x$ as coefficients to the Pauli spin matrices. $S$ has a global degree of
freedom under the adjoint representation of $\mathfrak{su}(2)$ so that we can
act on $S$ with an arbitrary matrix $U\in\mathrm{SU}(2)$ \begin{align} S
&\mapsto\tilde S= U^\dagger S U \end{align} maintaining the physically relevant
information. In particular, we can choose $U$ so that $S'$ becomes diagonal
\begin{align} \tilde S &=  \tilde x_3 \sigma_3\,,\label{eq:get_S_diagonal} \end{align} where
$\tilde x_3 = ||x||$. The rotation matrix $U$ leading to this result is not unique,
but for our purpose it is enough to find one particular realisation. Since we
measure the observable $S(\tau)$ at different values of the Euclidean time
$\tau$, we have to find one realisation $U_0$ that diagonalises $S(\tau_0)$ and
then apply the same transformation \begin{align} S(\tau) &\mapsto\tilde S(\tau) =
U_0^\dagger S(\tau) U_0^\pdagger\label{eq:transform_S_of_t} \end{align} for all
times.

It is essential to perform this rotation on every individual measurement before
averaging over the MC history.  Since there is no other gauge-fixing mechanism,
otherwise $S$ would average to 0.  In practice, the equal-time correlator for
$\tau=0$ is typically not representative for the other values of $\tau$.
Therefore, we used $\tau_0=1$ which has the maximally significant signal among
$\tau>0$.

From the relation between SU$(2)$ and SO$(3)$, or more intuitively the
equivalence between $S$ and $x$, it becomes clear that we can reduce the
diagonalisation problem to a rotation by $O\in\mathrm{SO}(3)$ with \begin{align}
Ox &=\tilde x\,.  \end{align} The diagonality condition~\eqref{eq:get_S_diagonal}
together with $x_3'>0$ then translates to \begin{align} \frac{x}{||x||} &=
O^\trans e_3\,.\label{eq:get_x_to_e3} \end{align} At this point we make the
ansatz to write $O$ as a product of planar rotations \begin{align} O &= R_\theta
R_\varphi\,,\\ R_\varphi &= \matr{ccc}{\cos\varphi & -\sin\varphi & 0 \\
\sin\varphi & \cos\varphi & 0 \\ 0 & 0 & 1}\,,\\ R_\theta &= \matr{ccc}{1 & 0 &
0 \\ 0 & \cos\theta & -\sin\theta \\ 0 & \sin\theta & \cos\theta}\,.
\end{align} The r.h.s.\ in equation~\eqref{eq:get_x_to_e3} then becomes
\begin{align} O^\trans e_3 &= \matr{c}{\sin\theta\sin\varphi \\
\sin\theta\cos\varphi \\ \cos\theta}\,.  \end{align} Setting this equal to the
l.h.s.\ of equation~\eqref{eq:get_x_to_e3} allows us to solve for $\varphi$ and
$\theta$ yielding \begin{align} \begin{split} \varphi &=
\atan\left(x_1,x_2\right)\,,\\ \theta &=
\atan\left(\sqrt{x_1^2+x_2^2},x_3\right)\,, \end{split}\label{eq:phi_and_theta}
\end{align} where $\atan$ denotes the 2-argument arctangent.  It is implemented
in most programming languages.  By definition, $\atan(y,x)$ returns the unique
$2\pi$-periodic angle between the coordinate $(x,y)$ and the positive $x$-axis.

Thus, the transformation~\eqref{eq:transform_S_of_t} of $S$ can be expressed by
the rotation \begin{align} x(\tau) &\mapsto\tilde x(\tau) = R_\theta R_\varphi
x(\tau) \end{align} where $\varphi$ and $\theta$ are determined by
equation~\eqref{eq:phi_and_theta} using $x(\tau_0)$. Simple sanity checks for
the correctness of the rotation are that $\tilde x(\tau_0)=(0,0,||x(\tau_0)||)^\trans$
and that $||\tilde x(\tau)||=||x(\tau)||$ for all $\tau$.

\printbibliography

\end{document}